\documentclass[final,1p,times,authoryear]{elsarticle}
\usepackage[utf8x]{inputenc}

\usepackage{amssymb}
\usepackage{multirow}
\usepackage[version=4]{mhchem}
\usepackage{arydshln}
\usepackage{subcaption}
\usepackage{multicol}
\usepackage[colorinlistoftodos]{todonotes}
\usepackage{lineno}

\usepackage{hyperref}
\hypersetup{
    colorlinks = false,
    linkcolor=red,  
    linkbordercolor = {1 0 1},
    citebordercolor={0 1 0},
}

\journal{Icarus}

\begin{document}
\begin{frontmatter}



\title{Positive Ion Chemistry in an N$_2$-CH$_4$ Plasma Discharge: Key Precursors to the Growth of Titan Tholins}

\author[1]{David Dubois\corref{corr}} 

\ead{david.f.dubois@nasa.gov}

\author[1]{Nathalie Carrasco} 

\author[1]{Lora Jovanovic} 

\author[1]{Ludovic Vettier}

\author[1]{Thomas Gautier}

\author[6]{Joseph Westlake}

\address[1]{LATMOS/IPSL, UVSQ Université Paris-Saclay, UPMC Univ. Paris 06, CNRS, Guyancourt, France}

\address[6]{Johns Hopkins University Applied Physics Laboratory, Laurel, MD, USA}

\cortext[corr]{Present addresses: NASA Ames Research Center, Space Science \& Astrobiology Division, Astrophysics Branch, Moffett Field, CA, USA, Bay Area Environmental Research Institute, Moffett Field, CA, USA.}



\begin{abstract}
Titan is unique
in the solar system as it hosts a dense atmosphere
mainly made of molecular nitrogen N$_2$ and methane
CH$_4$. The Cassini-Huygens Mission revealed the presence of an intense atmospheric photochemistry
initiated by the photo-dissociation and ionization of
N$_2$ and CH$_4$. 
In the upper atmosphere, Cassini detected signatures compatible
with the presence of heavily charged molecules
which are precursors for the solid core of the aerosols. These observations have indicated that ion chemistry
has an important role for organic growth. However,
the processes coupling ion chemistry and aerosol formation and growth are still mostly unknown. In this study, we investigated
the cation chemistry responsible for an efficient organic
growth that we observe in Titan's upper atmosphere,
simulated using the PAMPRE plasma reactor. Positive ion precursors were measured by \textit{in situ} ion mass
spectrometry in a cold plasma and compared with INMS observations taken during the T40 flyby. A series of positive ion measurements were performed in three \ce{CH4} mixing ratios (1\%, 5\% and 10\%) showing a variability in ion population. Low methane concentrations result in an abundance of amine cations such as \ce{NH4+} whereas aliphatic compounds dominate at higher methane concentrations. In conditions of favored tholin production, the presence of C$_2$ compounds such as \ce{HCNH+} and \ce{C2H5+} is found to be consistent with copolymeric growth structures seen in tholin material. The observed abundance of these two ions particularly in conditions with lower \ce{CH4} amounts is consistent with modeling work simulating aerosol growth in Titan's ionosphere, which includes mass exchange primarily between \ce{HCNH+} and \ce{C2H5+} and negatively charged particles. These results also confirm the prevalent role of C$_2$ cations as precursors to molecular growth and subsequent mass transfer to the charged aerosol particles as the \ce{CH4} abundance decreases towards lower altitudes.
\end{abstract}


\end{frontmatter}


\newpage
\section{Introduction}
\label{Introduction}

Titan is Saturn's largest satellite. It hosts a dense atmosphere mainly made up of molecular nitrogen N$_2$
and methane CH$_4$, with a surface pressure of 1.5 bar. Early seminal studies \citep{Khare1973,Sagan1973,Hanel1981} showed how Titan's reducing atmosphere
is also composed of complex and heavy organic molecules and aerosols. This chemistry is initiated at high altitudes and participates in the formation of solid organic particles \citep{Waite2007,Horst2017}. At these altitudes, the atmosphere is
under the influence of energy deposition such as solar ultraviolet (UV) radiation, solar X-rays, galactic cosmic rays, Saturn’s magnetospheric energetic electrons and solar wind \citep{Krasnopolsky2009,Krasnopolsky2014,Lavvas2011a,Sittler2009}. 

In Titan's upper atmosphere, Cassini’s Ion and Neutral Mass Spectrometer (INMS)
detected neutral and positive ion signatures \citep{Waite2007}. Subsequently, the
Cassini Plasma Spectrometer electron spectrometer (CAPS-ELS) unveiled the existence
of negative ion-molecules well over the detection range of INMS ($>$ 100 u)
consistent with the presence of heavy molecules (over 10,000 u in mass) which
are presumably precursors for the solid core of the aerosols \citep{Coates2007,Desai2017,Dubois2019b}. Furthermore, \cite{Lavvas2013} modeled the photochemistry and microphysics in the ionosphere, characterizing the interaction between the aerosols and charged particles. They showed the dusty nature of the ionosphere (where the charged aerosol population dominates the overall gas phase/aerosol particles charge balance), and found that the aerosol growth in the ionosphere, notably below 1,000 km, occurs as negatively charged particles collide with background positive ions. This in turn, leads to a rapid and important growth in mass (e.g. $\sim$ 500 u at 1,000 km). Ion-molecule reactions are thought to produce most of the positive ions present in Titan’s ionosphere, and are thus controlled by the two initial neutral main constituents, \ce{N2} and \ce{CH4}. The direct ionization of \ce{N2} and \ce{CH4} and formation of the \ce{N+} and \ce{N+2} primary ions make \ce{CH+3} readily available which is predicted to participate in the production of the first light hydrocarbons such as \ce{C2H5+}, one of the major ions produced (Reaction \ref{methane destruction}). The ionization rate peaks at the ionospheric peak, at about 1,150 km. \cite{Ip1990a} determined, using the Chapman layer theory, the electron density
peak to be at an altitude of 1,200 km, while \cite{Keller1992} found it at an altitude of 1,175 km. Later, \cite{Fox1997} predicted it to be at 1,040 km with a solar zenith angle of 60º. They also predicted that \ce{HCNH+} would be the major ion,
which included a model with over 60 species and 600 reactions. \cite{Keller1998}, with an improved model, estimated the major \textit{m/z} 28 peak to consist
of HCNH+ at 75\%. Some other major ion species include \ce{CH+5}, \ce{C3H+3} and \ce{C3H+5}. Based on \cite{Yung1984} and \cite{Toublanc1995}, \cite{Keller1998} predicted higher masses than previous models, to be detected by Cassini, of up to C$_6$ species in the ionospheric peak region.\\

\begin{equation}
\label{methane destruction}
\ce{CH_3^+ + CH_4 \longrightarrow C_2H_5^+ + H_2}
\end{equation}\\

Since the Cassini era, extended models based on \cite{Keller1998} have been used to predict ion densities and mass distributions, and construct a more cohesive view of neutral-ion interactions in Titan’s ionosphere. Photochemical models \citep[e.g.][]{Vuitton2007,Carrasco2008a,Vuitton2008} have
provided insight on the chemical species involved in ion-molecule, proton transfer
or ionization reactions to produce positive ions. In recent years, \textit{e.g.} \cite{Agren2009,Shebanits2013,Shebanits2017,Vigren2014,Vigren2015} have explored electron number
densities with Solar Zenith Angle (SZA) dependencies and dayside/nightside ion
charge densities with EUV flux correlations. This has shown how
the positive ion and dust grain charge densities are diurnally sensitive
to EUV fluctuations and thus impact ion density distributions.
Furthermore, the gas-to-solid conversion at these high altitudes coexists in a fully
coupled ionic and neutral chemistry \citep[e.g.][]{Lavvas2013,Vuitton2019}. However, the processes coupling ion and neutral
chemistry and aerosol production are still mostly unknown at the moment. Using laboratory simulations with different gas mixtures, it is possible to influence the chemistry and thus constrain species. In this way, laboratory results
can for example, provide potential new species, rule out some, or probe
masses with a higher resolution that were out of reach of the Cassini instruments,
while improving understanding of chemical pathways. Detecting these ions (positive or negative) in
Titan-like conditions in the laboratory remains challenging. \cite{Sciamma-OBrien2014} used Time-of-Flight mass spectrometry to study neutrals and positive ions produced in a pulsed plasma
jet expansion at low temperatures ($\sim$ 150 K). In such an apparatus, the ions are not fragmented by any kind of internal ionization. In that study, the authors focused on the first chemical stages before tholin (Titan aerosol analogs produced in the laboratory) production, by studying simple binary gas mixtures to more complex ones with light hydrocarbons and benzene. With comparisons to CAPS-IBS data, they suggested for example that some
of the ions of larger masses (\textit{m/z} $>$ 100) could correspond to aromatic compounds.\\

In the present study, we have used PAMPRE for the first time to study cations directly inside our radiofrequency capacitively coupled (RF-CC) plasma discharge with an ion mass spectrometer. We have used different \ce{N2}:\ce{CH4} initial gas mixtures ranging from 1\% to 10\% \ce{CH4} to address the complexity of the precursor gas phase chemistry which controls the complexity of tholins and pathway trends towards their formation. Finally, we have compared our experimental results with INMS measurements obtained during the T40 flyby.\\

\section{Materials and methods}

\begin{figure}
\centering
\includegraphics[width=1.0\textwidth]{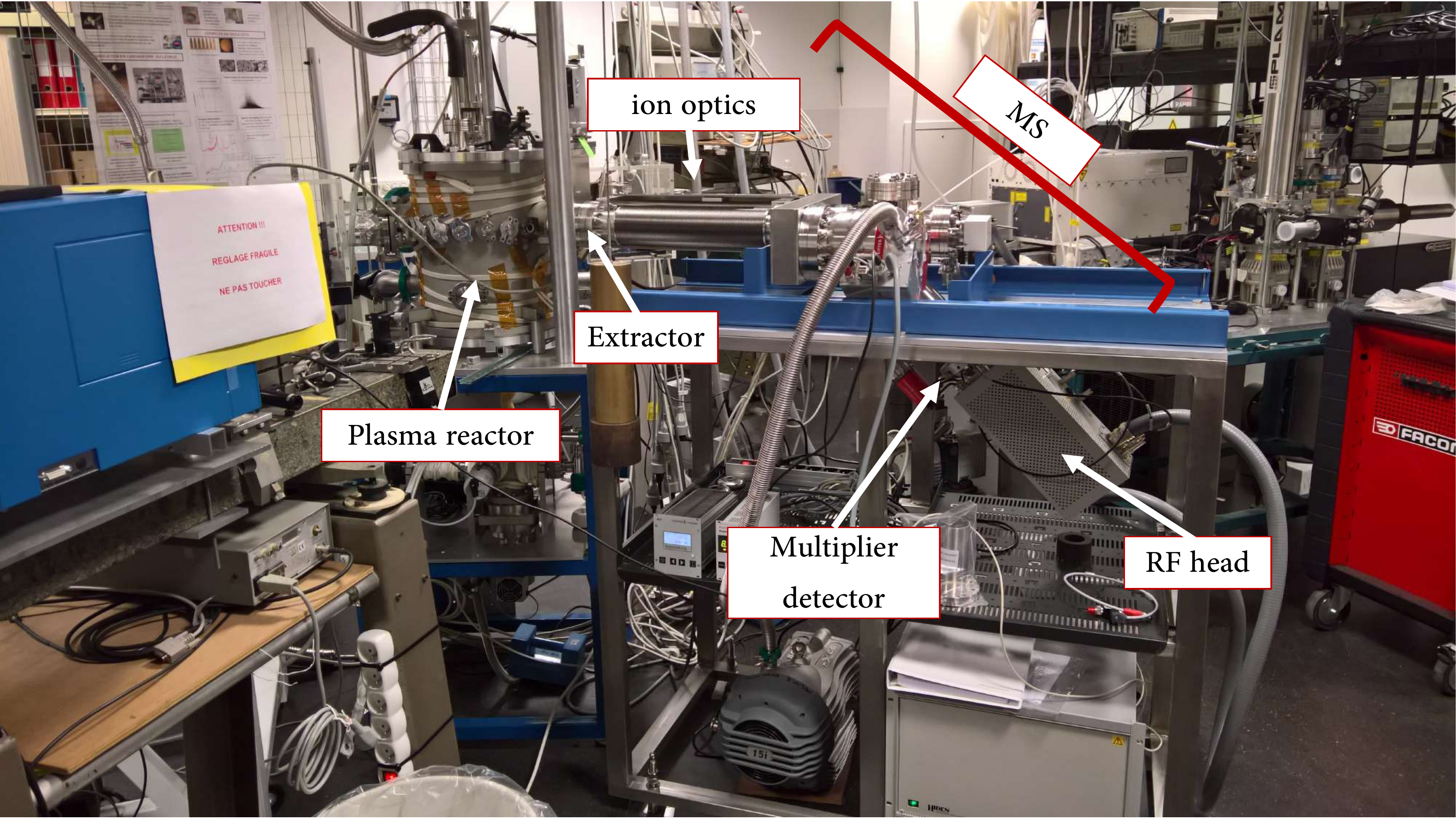}
\caption{\label{Fig1}The PAMPRE cold plasma chamber, along with its suite
of instruments. In particular, the ion and neutral mass spectrometer is
visible to the right of the chamber. The chamber and mass spectrometer
are separated by a VAT valve, enabling a residual gas pressure
within the transfer tube of 10$^{-9}$ mbar. The residual pressure in the
PAMPRE chamber is 10$^{-6}$ mbar.}
\end{figure}

We used the cold plasma reactor PAMPRE (\cite{Szopa2006} and Figure \ref{Fig1}) to reach analogous conditions to Titan's ionosphere using various initial \ce{CH4} conditions. This plasma reactor has previously been used to characterize the gas and solid phase particles produced in the discharge \citep[\textit{e.g.}][]{Carrasco2012,Gautier2012,Dubois2019b,Dubois2019a}. A newly acquired ion mass spectrometer was fitted to the existing chamber, in order to detect the ions directly within the plasma \textit{in situ}. Different experimental conditions were tested and are detailed hereafter. The plasma chamber and mass spectrometry techniques are presented in the following sections, respectively.

\subsection{The PAMPRE cold plasma experiment}

The PAMPRE plasma reactor delivers electrons with an energy distribution function
comparable to that of the solar spectrum \citep{Szopa2006}.
The plasma discharge is delivered through a radiofrequency (13.56 MHz) generator.
\ce{N2}:\ce{CH4} gas influx is monitored with mass flow controllers at 55 sccm (standard cubic centimeters per minute) in standard conditions, tuned from \ce{CH4} mixing ratios from 1\% to 10\%. This corresponds to a working pressure of $\sim$ 0.9 mbar. High purity ($>$ 99.999\%) cylinders of \ce{N2} and \ce{N2}:\ce{CH4} were used. Plasma power was kept constant throughout the experiment at 30 W.

\begin{figure}
\centering
\includegraphics[width=0.5\textwidth]{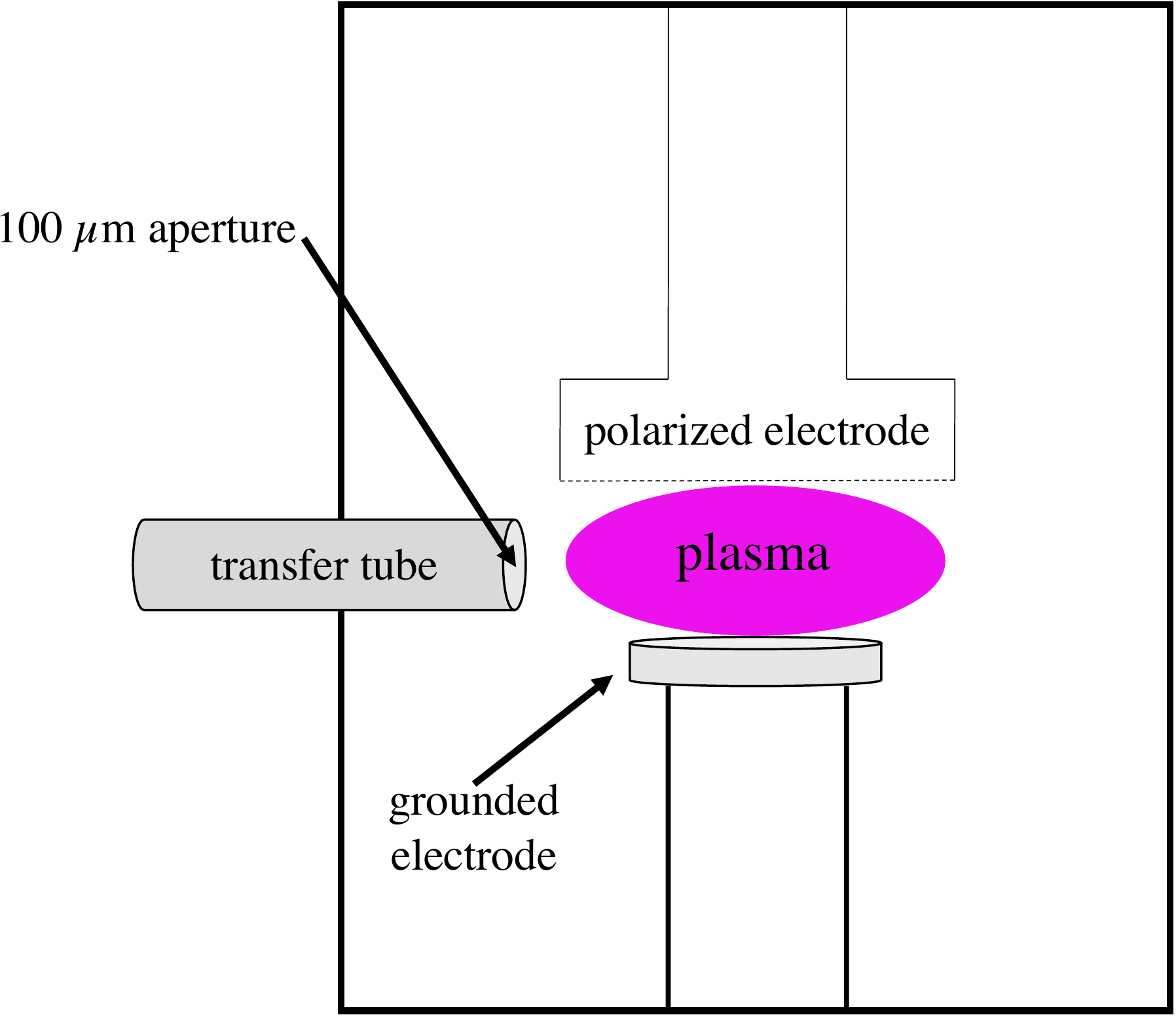}
\caption{\label{Fig2}Schematics of the mass spectrometer in contact with the plasma inside the reactor. The plasma is contained between the polarized and grounded electrodes. In this study we have used a 100 $\mu$m aperture, providing a reasonable balance between measurement integration time and signal count.}
\end{figure}

The plasma operates between a polarized electrode (anode) and a grounded electrode \citep{Szopa2006}.
The showerhead-shaped anode of PAMPRE can be configured in two different ways: with cage or cage-free. Previous studies using this reactor have analyzed the plasma parameters with a fitted cage \citep[e.g.][]{Wattieaux2015} ensuring a controlled geometry of the discharge, but preventing ion collection. Without this cage, an expanded plasma operates between the polarized electrode and a grounded plate at the bottom of the reactor, 10 cm from each other (Figure \ref{Fig2}). 

\subsection{Positive ion mass spectrometry}

\begin{figure}
\centering
\includegraphics[width=1.1\textwidth]{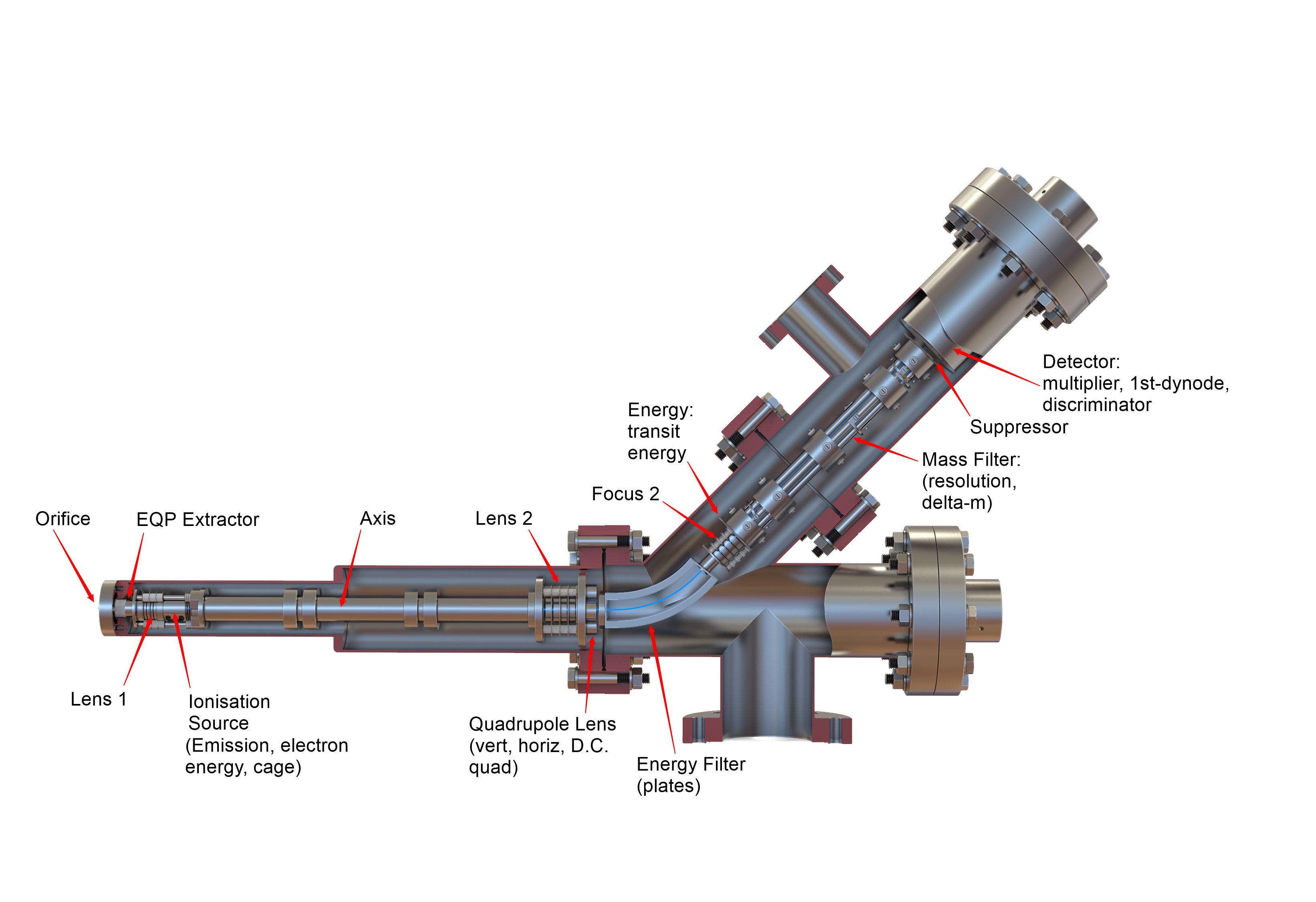}
\caption{\label{Fig3}Schematic diagram of the EQP by Hiden Analytics. The ions are
extracted from the plasma by the transfer tube on the left-hand side.
The extractor’s floating potential enables the extraction of the positive
ions. Then, the ions are guided by a series of lenses until they hit
the multiplier detector. We set the multiplier to 1800 V. The RF head
analyzes the impacted particles and sends the counts to a computer. Credit: Hiden Analytics}
\end{figure}

A new Hiden Analytical EQP 200 quadrupole mass spectrometer (Figure \ref{Fig3}), coupled with a Pfeiffer turbo pump, has been fitted to PAMPRE. Its
mass range goes up to 200 u. The turbo pump enables us to reach secondary
vacuum of 10$^{-9}$ mbar inside the MS chamber, and $\sim 4 \times 10^{-5}$ mbar during ion extraction. Different extraction apertures can be fitted to the extractor, depending on the pressure in the reactor. The extractor, where a negative potential is applied, directly extracts ions from the plasma. A gate valve was placed between the chamber and the spectrometer to enable the 10$^{-9}$ mbar vacuum inside the spectrometer and reduce the \ce{CO2} and \ce{H2O} contaminations as much as possible. However, an accumulation of tholins within the instrument is possible over time. To mitigate accumulation of tholins, which will contaminate the experiment, we both isolate the extractor and perform an ultrasound bath and/or an \ce{O2} plasma before and after each experiment.

A 0.1 u mass increment is used throughout the entire mass range for all
experiments. The ion extractor, in contact with the plasma, delivers ions from the plasma to the mass spectrometer for analysis (Figure \ref{Fig2}). This proved to be efficient
enough to have a high count detection ($\sim$ 10$^6$ c/s), while staying outside of the inter-electrode space.
The spectrometer unit converts signal into a number of counts-per-second, over an integration period
called hereafter \textit{dwell time}. The \textit{dwell time} is user-defined, and set to 100 ms. Once
enough signal has been accumulated, the measurement starts. With a 0.1 u mass step and a dwelling time of 100 ms over a 100 u range, a scan typically takes $(100 u) / 0.1 u \times 100 ms = 100$ s/scan. Any signal $<$ 10 c/s corresponds to instrumental and electrical background noise. Limiting air and water contamination in the plasma and spectrometer chambers is essential in avoiding over-interpreting spectra (\textit{e.g.}
masses 18 and 32). A motorized Z-drive is mounted to the ion extractor, giving a 300
mm stroke in order to approach the plasma closely ($<$ 2cm) and measure the ion
composition. This Z-drive system consists of a coil-shaped structure enclosing the
transfer tube and is prone to water condensation in the gap of each spiral. Water contamination can be evaluated from the contribution of its main ion in the discharge, which is \ce{H3O+} at \textit{m/z} 19. As seen on Figure \ref{fig:1-5-10pc-C1}, the peak at \textit{m/z} 19 is two orders of magnitude lower than the main ions in the C$_1$ block, confirming a weak water contamination in our experiment. Indeed before each experiment, we ensured an efficient degassing of water in the plasma reactor by a first bake-out of the transfer tube enclosure and reactor at 110$^{\circ}$C, followed by a pure \ce{N2} discharge removing residual adsorbed molecules on the walls.
Once the extractor has been inserted into the plasma, a scan is completed, satisfying both mass range detection and ion count intensity. The intensity
limit for this instrument is $\sim 10^7$ c/s. For further experimental details, the reader is referred to the Supplementary Material.

\section{Results}

We present here the positive ion mass spectra obtained in conditions ranging from [CH$_4$]$_0$ = 1$\%$ to [CH$_4$]$_0$ = 10$\%$ covering Titan methane conditions and staying consistent with previous studies \citep{Sciamma-OBrien2010,Dubois2019a}. Firstly, three [CH$_4$]$_0$ conditions are presented, showing the diversity in ion population. We used 1\%, 5\% and 10\% methane mixing ratios. Then, we show the disparity in ion distributions and compare our results with INMS measurements taken during the Cassini T40 flyby.

\subsection{Ion diversity}

\begin{figure}
\centering
\includegraphics[width=1.1\textwidth]{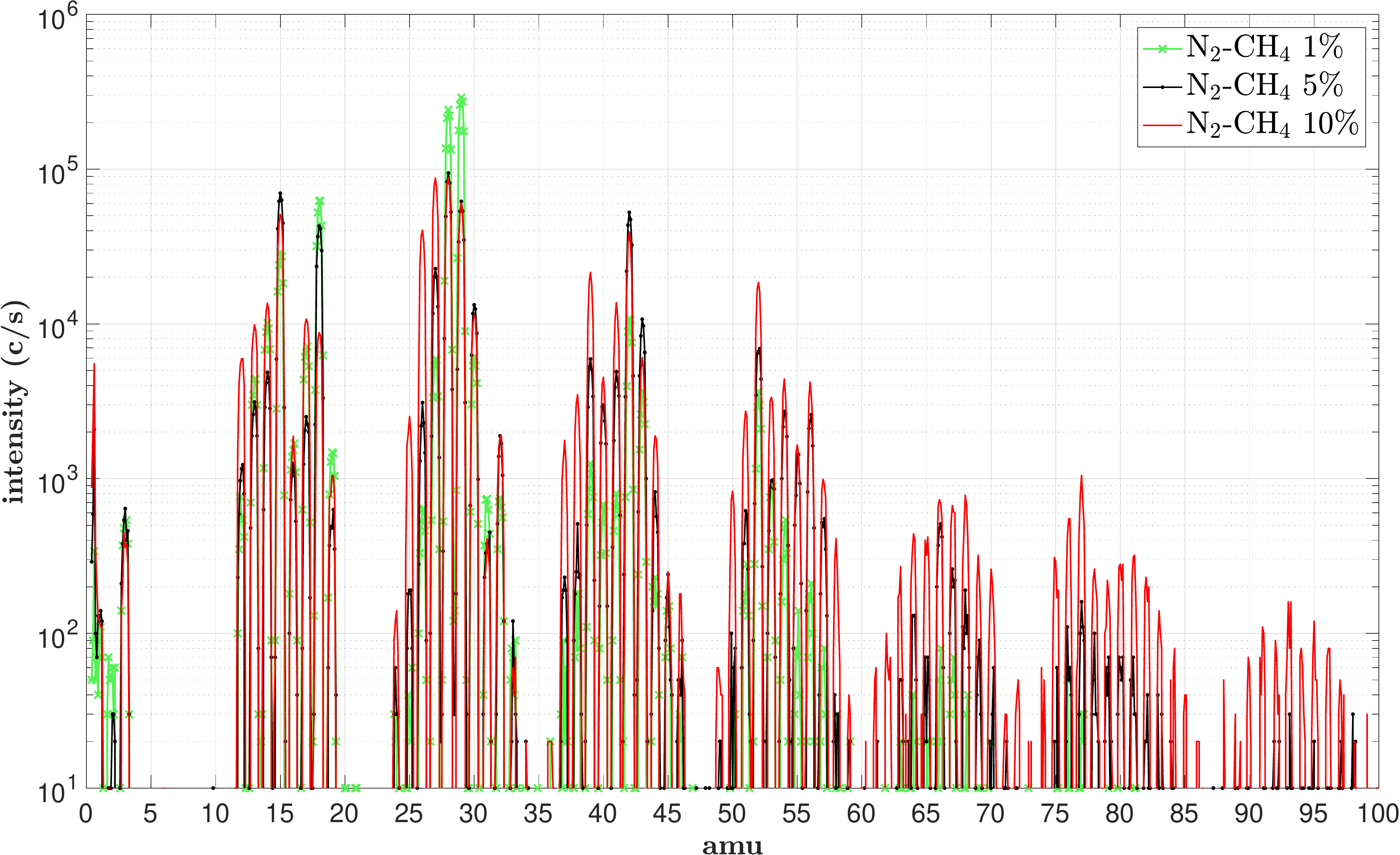}
\caption{\label{fig:1-5-10pc-energy-optimized}Mass spectra for [N$_2-$CH$_4$]$_0 = 1\%, 5\%$ and $10\%$ mixing ratio, in green, black and red, respectively. The energy filter was settled at 2.2, 2.2 and 1.2 V for the experiments at 1\%, 5\% and 10\%, respectively.}
\end{figure}

\begin{figure}
\centering
\includegraphics[width=1.0\textwidth]{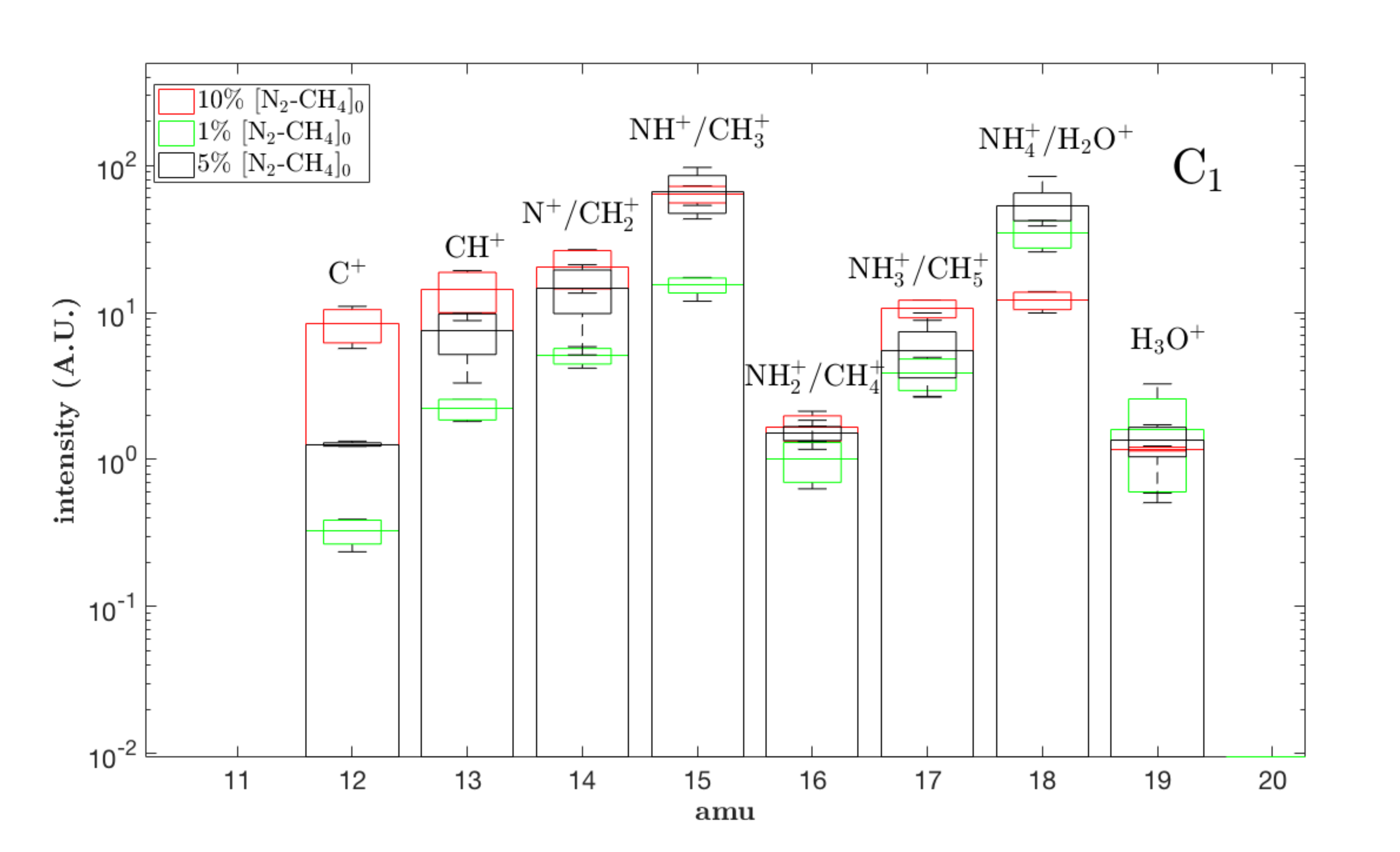}
\caption{\label{fig:1-5-10pc-C1}C$_1$ group in a [N$_2-$CH$_4$]$_0 = 1\%$ [N$_2-$CH$_4$]$_0 = 5\%$ and [N$_2-$CH$_4$]$_0 = 10\%$ mixing ratio.}
\end{figure}

\begin{figure}
\centering
\includegraphics[width=1.0\textwidth]{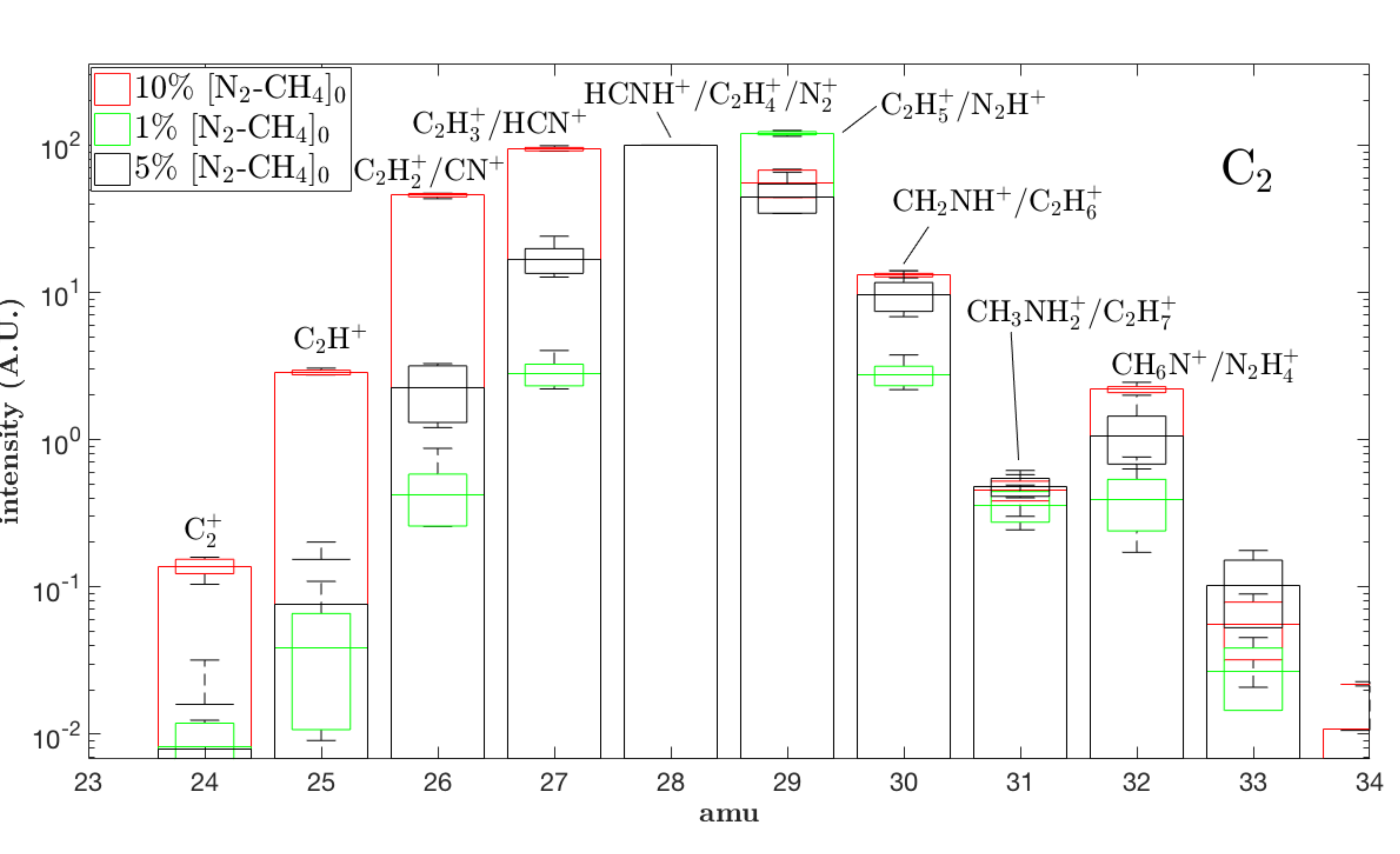}
\caption{\label{fig:1-5-10pc-C2}C$_2$ group in a [N$_2-$CH$_4$]$_0 = 1\%$ [N$_2-$CH$_4$]$_0 = 5\%$ and [N$_2-$CH$_4$]$_0 = 10\%$ mixing ratio.}
\end{figure}

\begin{figure}
\centering
\includegraphics[width=1.0\textwidth]{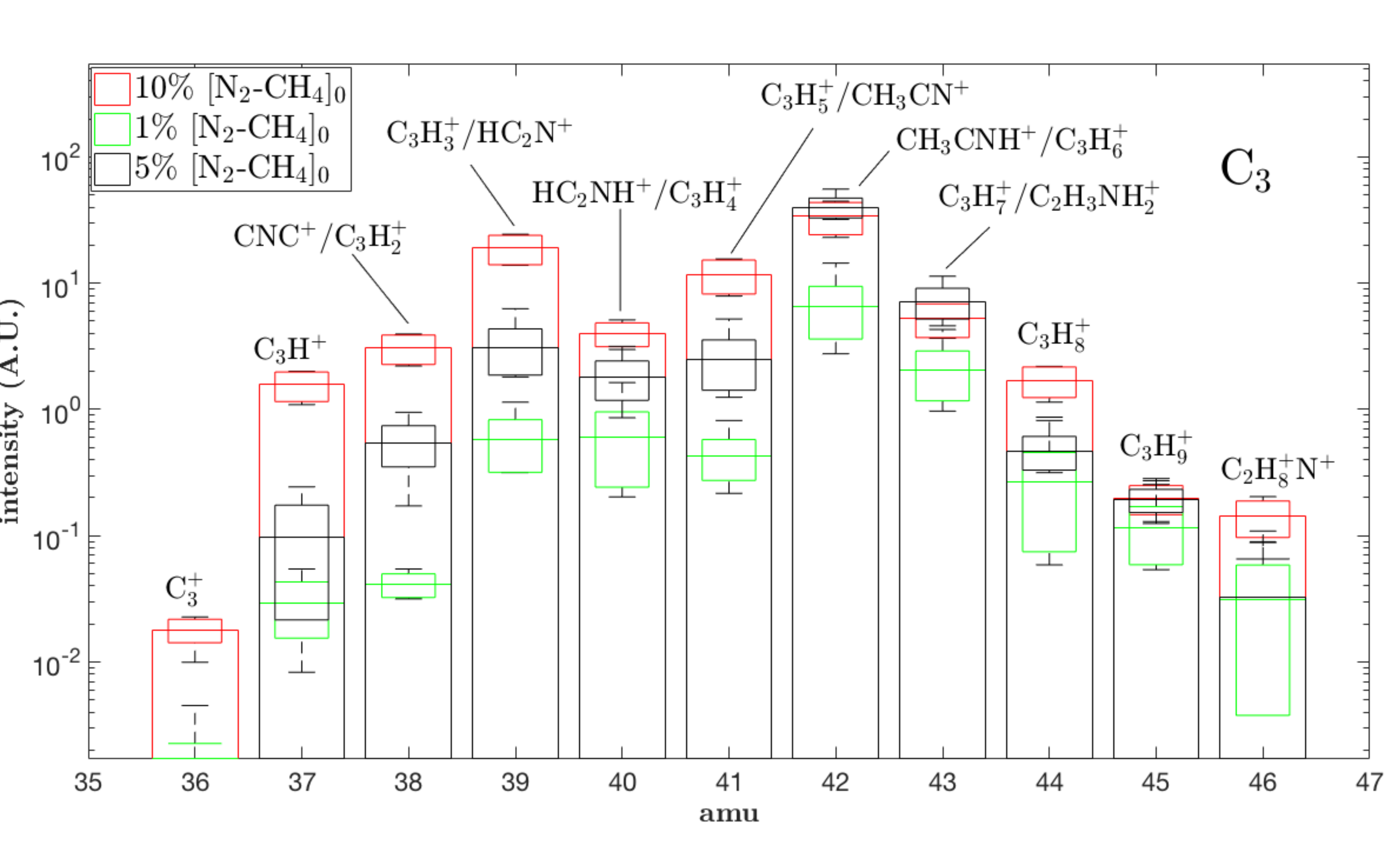}
\caption{\label{fig:1-5-10pc-C3}C$_3$ group in a [N$_2-$CH$_4$]$_0 = 1\%$ [N$_2-$CH$_4$]$_0 = 5\%$ and [N$_2-$CH$_4$]$_0 = 10\%$ mixing ratio.}
\end{figure}

\begin{figure}
\centering
\includegraphics[width=1.0\textwidth]{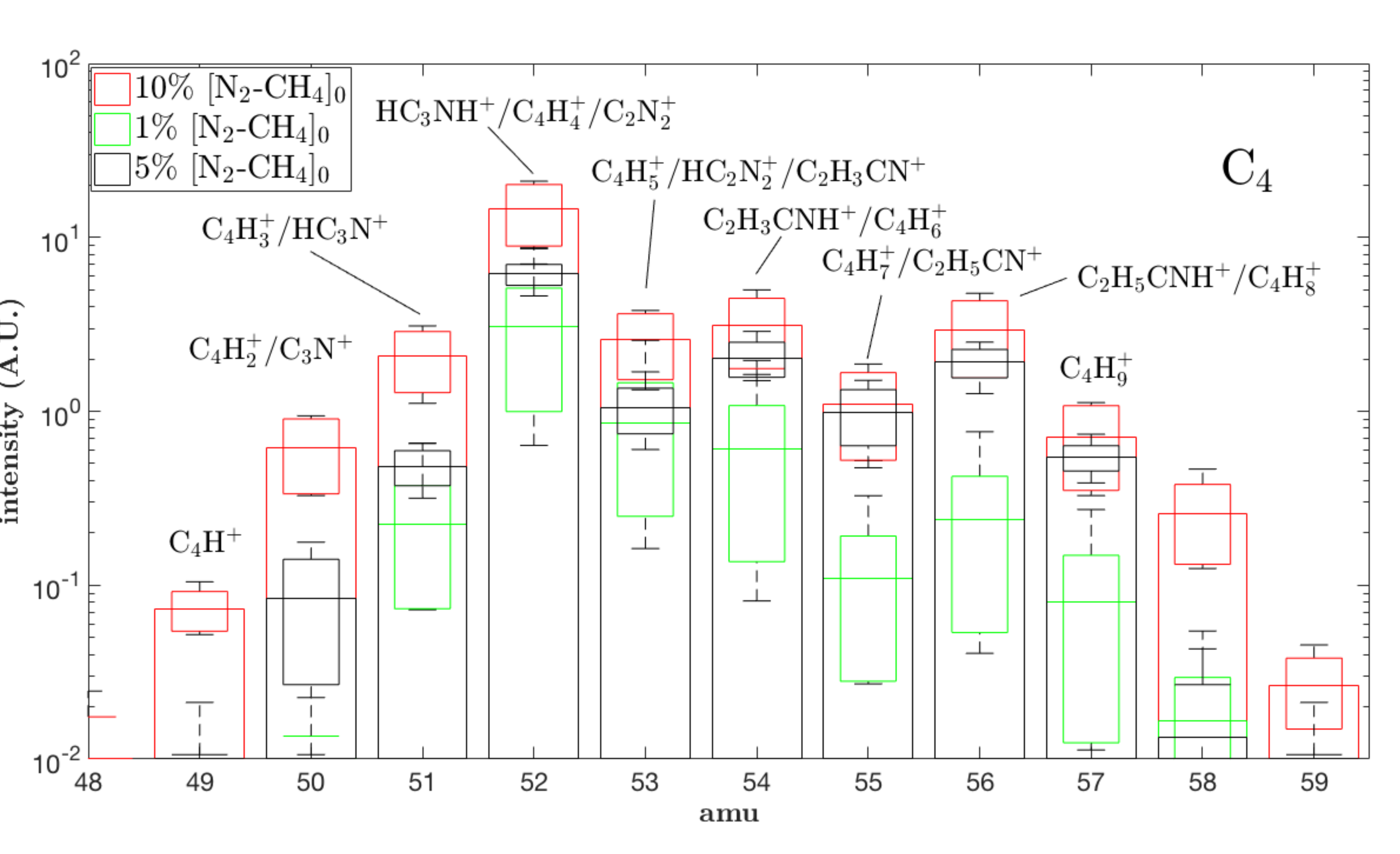}
\caption{\label{fig:1-5-10pc-C4}C$_4$ group in a [N$_2-$CH$_4$]$_0 = 1\%$ [N$_2-$CH$_4$]$_0 = 5\%$ and [N$_2-$CH$_4$]$_0 = 10\%$ mixing ratio.}
\end{figure}

As explained in the Supplementary Material, all spectra are normalized to \textit{m/z} 28 and units are expressed in arbitrary units (A.U.). Figure \ref{fig:1-5-10pc-energy-optimized} shows a large variability (e.g. C$_8$ blocks) in the spectra.
Here, we tentatively assign positively charged species to the measured peaks. These
assignations are listed in Table \ref{table 3}. We focus on the first four groups (\textit{m/z} $<55$) as it becomes speculative to attribute higher masses beyond this limit (influence of
polymers, aromatics...). These group comparisons for three different initial methane
mixing ratios are shown in the box plots Figures \ref{fig:1-5-10pc-C1} through \ref{fig:1-5-10pc-C4}. Plotted values correspond to averaged normalized values taken from four data sets in each initial methane concentration. Lower and upper limits of the boxes represent the 1st and 3rd quartiles, respectively, of the minima and maxima experimental values (black whiskers), indicating the variability in intensity displayed in all four spectra. The spectra in green, black and red correspond to measurements carried out with 1\%, 5\% and 10\% \ce{CH4}, respectively.

\begin{table}
	\centering
	\caption{Tentative attributions of several species from the C$_1$, C$_2$, C$_3$ and C$_4$ groups. These attributions are based on INMS observations and model-dependent calculated ion densities from \cite{Vuitton2007}. We give the presumed ions possible for each mass. Water ions are also included, the main one being \ce{H3O+} at \textit{m/z} 19. Its signal is two orders of magnitude lower than  the rest of the other ions in the C$_1$ block, confirming the weak water contamination in the experiment (see section 2.2).}
	\label{table 3}
	\begin{tabular}{ccc} 
		\hline
		Group & \textit{m/z} & Species\\
		\hline
        \multirow{6}{*}{C$_1$} &
        14 & N$^+$/CH$_2^+$\\
		& 15 & CH$_3^+$/NH$^+$\\
		& 16 & CH$_4^+$/NH$_2^+$\\
		& 17 & CH$_5^+$/NH$_3^+$\\
        & 18 & NH$_4^+$/H$_2$O$^+$\\
        & 19 & H$_3$O$^+$\\
        \hline
        \multirow{5}{*}{C$_2$} &
        26 & C$_2$H$_2^+$/CN$^+$\\
        & 27 & C$_2$H$_3^+$/HCN$^+$\\
        & 28 & HCNH$^+$/N$_2^+$/C$_2$H$_4^+$\\
        & 29 & C$_2$H$_5^+$/N$_2$H$^+$/CH$_2$NH$^+$\\
        & 30 & CH$_2$NH$_2^+$/C$_2$H$_6^+$\\
        \hline
        \multirow{8}{*}{C$_3$} &
        38 & CNC$^+$/C$_3$H$_2^+$\\
        & 39 & C$_3$H$_3^+$/HC$_2$N$^+$\\
        & 40 & HC$_2$NH$^+$/C$_3$H$_4^+$\\
        & 41 & CH$_3$CN$^+$/C$_3$H$_5^+$\\
        & 42 & CH$_3$CNH$^+$/N$_3^+$/C$_3$H$_6^+$\\
        & 43 & C$_3$H$_7^+$/C$_2$H$_3$NH$_2^+$\\
        & 44 & C$_3$H$_8^+$/CO$_2^+$\\
        & 45 & C$_3$H$_9^+$\\
        \hline
        \multirow{7}{*}{C$_4$} &
        51 & HC$_3$N$^+$/C$_4$H$_3^+$\\
        & 52 & HC$_3$NH$^+$/C$_2$N$_2^+$/C$_4$H$_4^+$\\
        & 53 & C$_4$H$_5^+$/HC$_2$N$_2^+$/C$_2$H$_3$CN$^+$ and \textit{m/z} 52 isotopes\\
        & 54 & C$_2$H$_3$CNH$^+$/C$_4$H$_6^+$\\
        & 55 & C$_4$H$_7^+$/C$_2$H$_5$CN$^+$\\
        & 56 & C$_4$H$_8^+$/C$_2$H$_5$CNH$^+$\\
        & 57 & C$_4$H$_9^+$\\
		\hline
	\end{tabular}
\end{table}

\begin{itemize}
\item $C_1$ - In Figure \ref{fig:1-5-10pc-C1}, the major ion detected at 5\% and 10\% CH$_4$ is at \textit{m/z} 15, which can be attributed to the \ce{NH+} radical or the methyl CH$_3^+$ cation. Other CH$_4$ fragments, consistent with a \ce{CH4} fragmentation pattern are also visible in all conditions at \textit{m/z} 14 and 13, corresponding to CH$_2^+$ and CH$^+$. The N$^+$ fragment also contributes to \textit{m/z} 14. CH$_5^+$ and  NH$_3^+$ may contribute to \textit{m/z} 17. There is a strong discrepancy between the 1\% and 10\% conditions, in particular regarding the \textit{m/z} 15 and \textit{m/z} 18 peak inversions. In a [N$_2-$CH$_4$]$_0 = 10\%$ mixing ratio, \textit{m/z} 15 dominates the rest of the grouping, presumably due to the methane influence along with \ce{CH+}, \ce{CH2+} and \ce{CH3+}. \citet{Carrasco2012} discussed the influence between the methyl ion and the favored production of ammonia with increasing methane. \cite{Mutsukura2001}, also using a CH$_4$/N$_2$ RF plasma albeit with much higher methane concentrations ($>$50\%), found that the \textit{m/z} 15 and \textit{m/z} 18 ions were dominant in methane-rich conditions. This is confirmed by our results (Figure \ref{fig:1-5-10pc-C1}), showing the importance of CH$_3^+$ in methane-rich mixtures. However, in a nitrogen-rich plasma with a mixing ratio of [N$_2-$CH$_4$]$_0 = 1\%$, \textit{m/z} 18 NH$_4^+$ becomes the most abundant ion. At 5\% \ce{CH4}, \textit{m/z} 18 is also important along with \textit{m/z} 15. Water contamination cannot be ruled out and is included in Table \ref{table 3}.
\end{itemize}

\begin{itemize}
\item $C_2$ - Figure \ref{fig:1-5-10pc-C2} shows the C$_2$ species detected in the three conditions. This group is overall more intense than the $C_1$ group. The most abundant ions are at \textit{m/z} 28 and 29 with intensities of $3 \times 10^1-10^2$ a.u. There is a strong increase in masses on the left tail of the red distribution, i.e. with 10\% \ce{CH4} (\textit{m/z} 24--27), as compared to the other two spectra. These species consistently increase well apart from the repeatability error bars with increasing methane. Presumably, these simple aliphatics such as C$_2$H$_2^+$ and C$_2$H$_3^+$ (\textit{m/z} 26 and 27, respectively) increase by an order of magnitude from the black to red curve. \textit{m/z} 29  is not as unambiguous, and \ce{N2H+} shares the same peak as C$_2$H$_5^+$. This peak is the most intense at 1\% \ce{CH4}. As such, where \textit{m/z} 28 dominates with a 10\% \ce{CH4} mixing ratio, the \textit{m/z} 29 peak inversion occurs with decreasing CH$_4$. Therefore, N-bearing species such as N$_2$H$^+$/\ce{CH2NH+} are good candidates at this mass. Due to the limiting amount of \ce{CH4} in the green spectrum, the ion at \textit{m/z} 29 in this condition is likely to be \ce{N2H+}. However, with higher methane amounts, \ce{C2H5+} formation prevails and becomes the dominant ion at \textit{m/z} 29, as predicted by \citet{Vuitton2007}. \ce{C2H5+} formation partly depends on the presence of \ce{C2H4}, indicating its presence in the plasma, as suspected in \citet{Dubois2019a}. \ce{CH2NH2+/C2H6+} at \textit{m/z} 30 increases with higher methane concentrations. It can thus most likely be attributed to \ce{C2H6+}.
\end{itemize}

\begin{itemize}
\item $C_3$ - This cluster is dominated by \textit{m/z} 42 (Figure \ref{fig:1-5-10pc-C3}), whose overall intensity seems to be favored with increasing methane. The black and red curves reach intensities similar to the $C_1$ group (almost $10^2$). Except in the 10\% CH$_4$ condition, the second most abundant volatile is at \textit{m/z} 43, which can be attributed to the protonated form of propene, C$_3$H$_7^+$, or ethylenimine \ce{C2H3NH2+}. With 10\% CH$_4$, the intensity of the \textit{m/z} 39 peak, attributed to C$_3$H$_3^+$ or \ce{HC2N+}, increases by almost two orders of magnitude. The production of this ion seems to be favored with a richer methane mixing ratio. Similar to the sharp increase seen in Figure \ref{fig:1-5-10pc-C2} (red bars) in signal on the left-half of the C$_2$ block, an important contribution of species present between \textit{m/z} 36 and \textit{m/z} 39 accompanies higher methane concentrations. This is consistent with proton addition reaction from hydrocarbon compounds.
\end{itemize}

\begin{itemize}
\item $C_4$ - This block is dominated by the ion at \textit{m/z} 52, which can correspond to HC$_3$NH$^+$, \ce{C4H4+} or \ce{C2H3CN+} (Figure \ref{fig:1-5-10pc-C4}). We observe an overall increase in intensity of these species with increasing methane. Similar to the C$_3$ case, this increase is especially relevant for the species between \textit{m/z} 49 and 52 in the 10\% \ce{CH4} condition. In addition, the increase in intensity at \textit{m/z} 52 with 10\% \ce{CH4} suggests that \ce{C4H4+} contributes significantly at this mass. Ions produced in the plasma with a 1\% \ce{CH4} mixing ratio remain overall less intense than in the other two \ce{CH4} conditions. To better characterize this observation among the different blocks, we represent these general trends as pie charts (Figure \ref{Pie charts 1,5,10pc}) as a function of initial methane concentration.
\end{itemize}

Each pie chart is color coded (horizontally and vertically) and is normalized over the sum of the mean values of all of the other masses (noted \textit{misc.} in gray color). This representation helps to draw the relative weights of each C$_x$ group and better visualize their relative contributions. Vertical color coding is represented according to the delta values of each \textit{m/z} using the ion series analysis technique \citep{McLafferty1993,Canagaratna2007}. Each \textit{m/z} is assigned a $\Delta$ value, where $\Delta = m/z - 14n +1$ and $n$ is the grouping number. Thus, $\Delta = 2$ corresponds to linear saturated hydrocarbons (light blue), $\Delta > 2$ corresponds to heteroatomic compounds (yellow, green and brown) and $\Delta < 2$ corresponds to unsaturated and branched hydrocarbons (dark blue, aquamarine, red, orange and purple). Therefore, the $\Delta$ value informs on the probable functional groups for each mass (the reader is referred to Figure 5 of the Supplementary Material for further information on the delta value determination).

The C$_1$ species are dominated by the peak at \textit{m/z} 18, which represents $\sim$ 11\% normalized over the entire spectrum in the 1\% \ce{CH4} condition, $\sim$ 13\% at 5\% \ce{CH4} and decreases to $\sim$ 2\% with 10\% \ce{CH4}. As we can see in the last condition, ions at \textit{m/z} 15 accounts for the same contribution as \textit{m/z} 18 does at 1\% \ce{CH4}. In the intermediate methane condition, both ions at \textit{m/z} 15 and \textit{m/z} 18 can be considered major ions. The C$_2$ compounds in a 1\% \ce{CH4} mixing ratio are significant, and represent almost 75\% in intensity of the entire spectrum. This group is mainly represented by the ions at \textit{m/z} 28 and \textit{m/z} 29. This C$_2$ influence decreases slightly at 5\% \ce{CH4} while still contributing to $\sim$ 44\% of all species detected. We observe that with increasing initial methane, the C$_2$ distribution gets wider and more populated, with eventually a more intense signature of peaks at \textit{m/z} 26, \textit{m/z} 29 and \textit{m/z} 30, at 8\%, 10\% and 2\%, respectively. This pattern is also seen in the C$_3$ pie charts, and to a lesser extent, the C$_4$ group. The C$_3$ compounds are dominated by ions at \textit{m/z} 42 in all three conditions. It is significant in particular at 5\% \ce{CH4} ($\sim$ 10\% normalized), a favorable condition for efficient tholin formation. As seen in Table \ref{table 3}, the peak at \textit{m/z} 42 can either be attributed to protonated acetonitrile \ce{CH3CNH+} or the cyclopropane positive ion \ce{C3H6+}.


\begin{figure}
\centering
\includegraphics[width=1.2\textwidth]{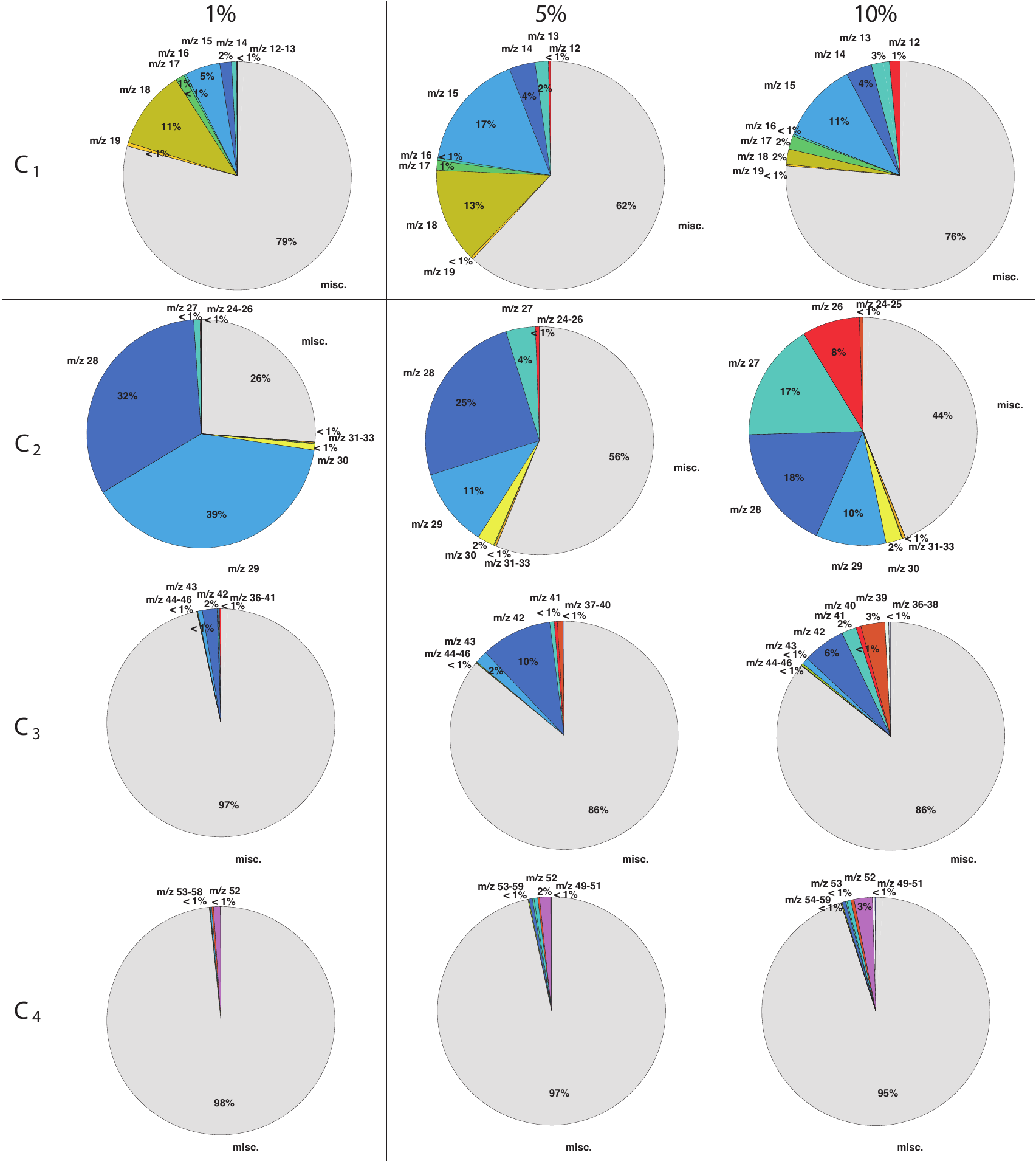}
\caption{\label{Pie charts 1,5,10pc}Pie charts representing the normalized mean intensities, for the C$_{1-4}$ molecular groups, as a function of the methane initial concentration (\%). Each chart is normalized by the total sum of all mean normalized intensities detected from 0-100 amu. Colored slices are the relative contributions in intensities of each respective peak over the summed intensities of all peaks of the spectrum. The rest of the peaks (1-C$_{1-4}$) are categorized in the gray slices (noted \textit{misc.}). Vertical color coding is based off ion series delta values (the readear is referred to Figure 5 of the Supplementary Material for further information).}
\end{figure}

It is noteworthy to point out the 5\% initial methane concentration charts (middle column, Figure \ref{Pie charts 1,5,10pc}). At this concentration, tholins are efficiently produced in our chamber. The presence of the C$_1$ (at \textit{m/z} 15 and 18) and C$_2$ (at \textit{m/z} 28 and 29) ions in this condition (\ce{CH3+}/\ce{NH4+} and \ce{HCNH+}/\ce{C2H5+}, respectively) is significant. This comes in agreement with the copolymeric HCN/\ce{C2H4} and poly-(\ce{CH2})$_m$(HCN)$_n$-based structure found in the solid tholin material \citep{Pernot2010,Gautier2014,Maillard2018}. As the tholins are negatively charged, they may react with these major ions to nucleate and polymerize. These results could suggest that \ce{HCNH+} and \ce{C2H5+} are important contributors to the polymeric growth of tholins.

\begin{figure}
\centering
\includegraphics[width=1.2\textwidth]{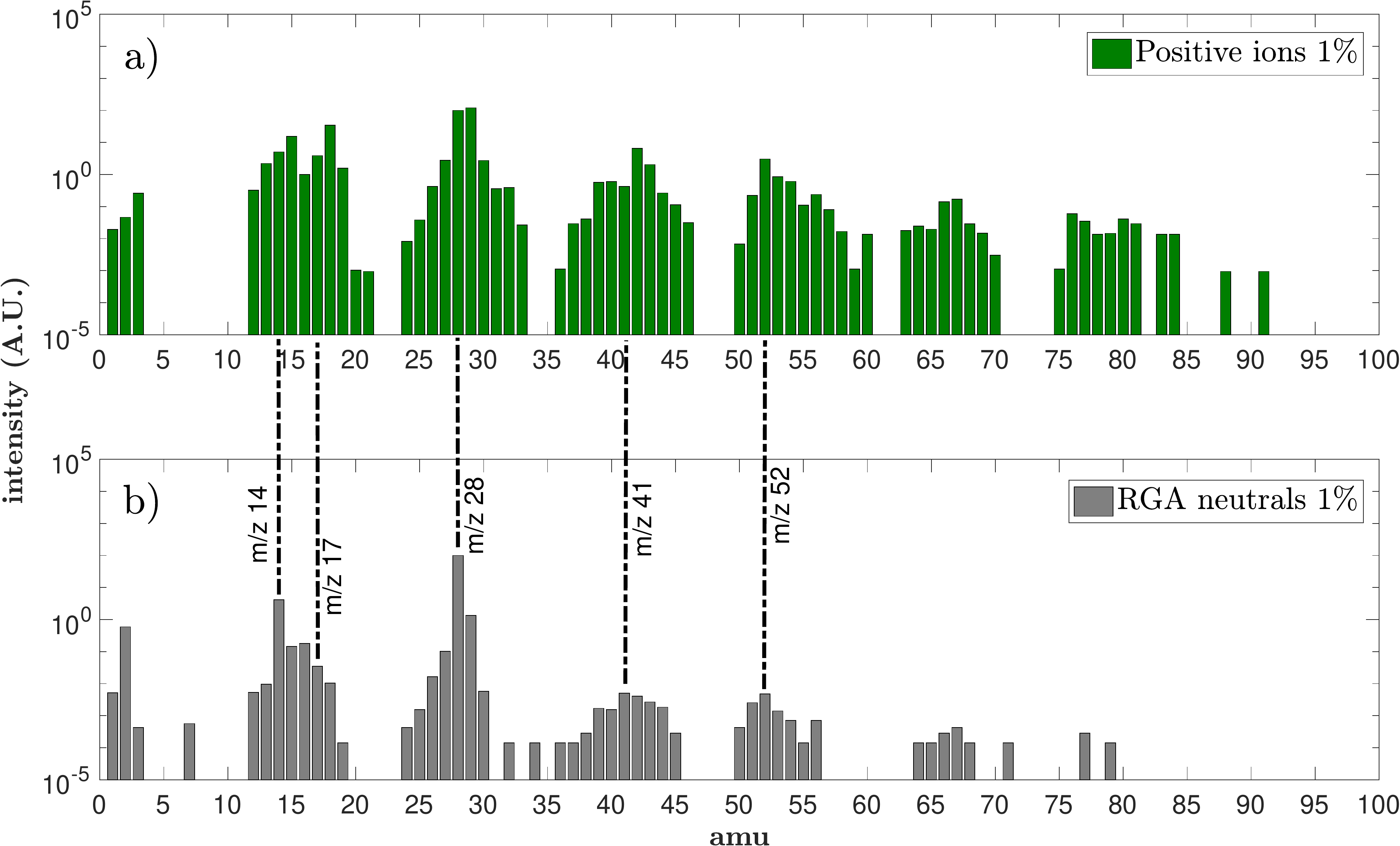}
\caption{\label{fig:neutrals-ions-comparison-1pc}Comparison of two normalized mass spectra taken with 1\% \ce{CH4} and degraded at the same resolution of 1 amu. a) shows the entire normalized averaged spectrum in positive ion mode of Figures \ref{fig:1-5-10pc-C1}--\ref{fig:1-5-10pc-C4}, while b) was taken with the plasma discharge on in RGA neutral mode, with an electron energy of 70 V and filament emission of 5 $\mu$A in the same conditions.}
\end{figure}

\begin{figure}
\centering
\includegraphics[width=1.2\textwidth]{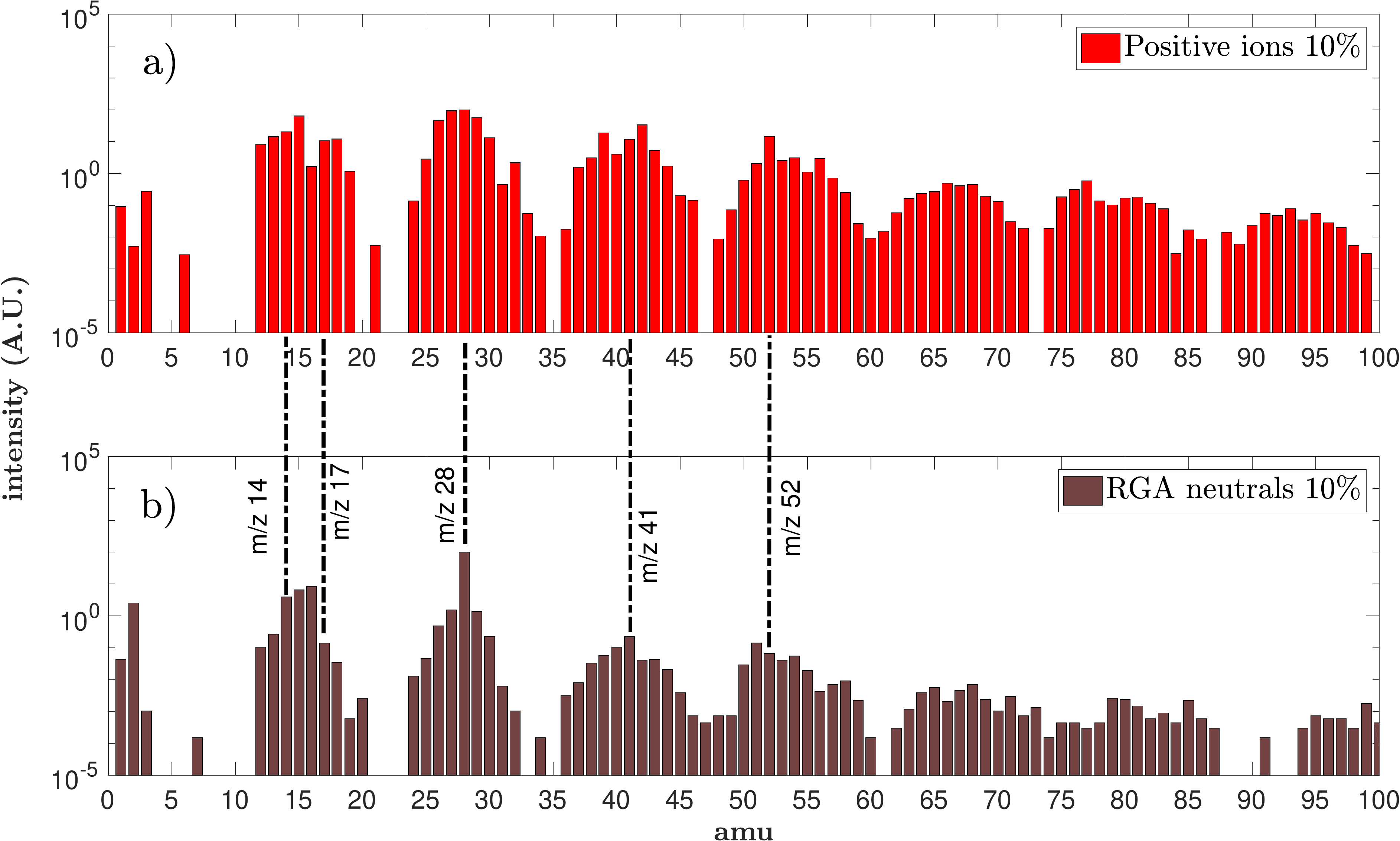}
\caption{\label{fig:neutrals-ions-comparison-10pc}Same as Figure \ref{fig:neutrals-ions-comparison-1pc}, with 10\% \ce{CH4}. Comparison of two normalized mass spectra taken with 10\% \ce{CH4} and degraded at the same resolution of 1 amu. a) shows the entire normalized averaged spectrum in positive ion mode of Figures \ref{fig:1-5-10pc-C1}--\ref{fig:1-5-10pc-C4}, b) the spectrum was taken in RGA neutral mode with the plasma discharge on, with an electron energy of 70 V and filament emission of 5 $\mu$A under the same conditions.}
\end{figure}

\subsection{Comparison between neutral and cation species}

In order to see the interactions between the neutral and positive ion species, we show in Figures \ref{fig:neutrals-ions-comparison-1pc} and \ref{fig:neutrals-ions-comparison-10pc} comparisons between the averaged normalized spectra in positive ion and residual gas analyzer (RGA) neutral modes at 1\% and 10\% \ce{CH4}, respectively. The neutral mode setting used a 70 V electron energy using a filament emission of 5 $\mu$A. \cite{Dubois2019a} had previously studied neutral volatile products released after being cryotrapped in the chamber, combining mass spectrometry and IR analysis for identification of major neutral products.

\begin{itemize}

\item  The ions at \textit{m/z} 28 and 29 in Figure \ref{fig:neutrals-ions-comparison-1pc} (a), dominate the spectrum as seen previously in Figure \ref{Pie charts 1,5,10pc}. In the neutral spectrum (gray bar graph), \textit{m/z} 28 corresponds to contributions from \ce{N2} and \ce{C2H4}. In the cryogenic study \citep{Dubois2019a}, \textit{m/z} 28 was attributed to \ce{C2H4}, while the major \textit{m/z} 17 peak was assigned to \ce{NH3}. In the middle plot, the ions at \textit{m/z} 14, corresponding to the nitrogen atom or methylene \ce{CH2}, is the main C$_1$ fragment. In positive ion mode however, the C$_1$ group is much more populated, as are all of the other groups. In addition, the ions at \textit{m/z} 15 and \textit{m/z} 18 become the two major ions, coupled with a relatively richer spectrum on the right hand side of the group compared with the RGA spectrum. This 1 u shift is consistent with reactions involving \ce{H+} addition (see Discussion Section). This pattern is also seen in the C$_2$ and C$_3$ groups with \textit{m/z} 28 to 29 and \textit{m/z} 41 to 42 shifts. The C$_4$ shows no inversions in the \textit{m/z} 52 peak. This molecule could be 1-buten-3-yne \ce{C4H4} or cyanogen \ce{C2N2}, which were both identified in \cite{Gautier2011}. Their cation counterparts \ce{C4H4+} and \ce{C2N2+}, along with propiolonitrile \ce{HC3NH+} could explain the intense \textit{m/z} 52 signature at $\sim 2.5$ a.u.

\item As seen at the beginning of this section and in Figure \ref{Pie charts 1,5,10pc}, the 10\% \ce{CH4} is marked by broader blocks, notably for the first three. The ions at \textit{m/z} 12 to 16 of methane fragments in the (b) plot of Figure \ref{fig:neutrals-ions-comparison-10pc} are visible along with the corresponding methylene and methyl cations (red plot). C$_2$ species are dominated by the peak at \textit{m/z} 28 surrounded by \textit{m/z} 26, 27 and 29. In our cryogenic study \citep{Dubois2019a}, the peak at \textit{m/z} 28 was attributed to \ce{C2H4}. \ce{N2} was absent from the chamber and so was ruled out. In the middle plot (b), \ce{N2} was also present in the plasma and can also contribute to this peak. Furthermore, the cryogenic analysis in \cite{Dubois2019a} showed a higher HCN production at 10\% than at 1\% \ce{CH4}, indicating that HCN must significantly contribute to the C$_2$ group with 10\% CH$_4$. There is a mass shift between the most intense peaks of the C$_3$ group, i.e. from \textit{m/z} 41 to 42 (in the brown and red plots, respectively), similar to that at 1\% \ce{CH4} (gray and green spectra, respectively), and again from \textit{m/z} 51 in the neutral spectrum to \textit{m/z} 52 only seen at 10\% \ce{CH4}. However, unlike in the 1\% \ce{CH4} condition, the increase in intensity of the peak at \textit{m/z} 42 is accompanied by the increase in intensity of the \textit{m/z} 39 peak, which represent 6\% and 3\% in relative intensity to the rest of the spectrum, respectively (Figure \ref{Pie charts 1,5,10pc}). To a lesser extent, this is also visible with the increase in intensity at \textit{m/z} 39 relative to \textit{m/z} 41 from the neutral (Figure \ref{fig:neutrals-ions-comparison-1pc}b) to the ion spectra (Figure \ref{fig:neutrals-ions-comparison-1pc}a) at 1\% \ce{CH4}.
\end{itemize}

This comparison between spectra taken in neutral and positive ion mode shows discrepancies and mass shifts between both datasets. Notably, mass shifts and sometimes peak inversions are visible in the first four C$_x$ groups, which could be due to an enrichment of hydrogenated ion species, indicative of different chemistry. To go further in this inter-comparison and better understand specific chemical processes, future work will focus on modeling the neutral/cation interaction in the plasma.

\section{Discussion}
\subsection{Chemical pathways contributing to the major precursors}\
\label{Chemical pathways}

The abundant reservoir of neutral species in Titan's upper atmosphere is an important source for ion-molecule reactions. Ion chemistry is initialized by dissociative and ionizing processes of major neutral compounds forming \ce{N2+} and \ce{N+}, and \ce{CH4+}, \ce{CH3+}, \ce{CH2+} and \ce{CH+} (see Introduction). These simple and primary cations are essential for further ion-neutral reactions (Reaction \ref{methane destruction}). 
Thanks to photochemical models (see Introduction and references therein), many of the positive ions have been shown to be products resulting from proton exchange reactions between neutrals and ions. In this section, we will discuss potential pathways to explain some of our main ions detected in our three methane mixing ratios. These pathways are detailed in order to constrain our understanding of the cation gas phase reactivity in our plasma.\\

The ion variability observed in our PAMPRE reactor has shown to be highly methane-dependent. Few peak inversions exist (e.g. \textit{m/z} 15 and 18). However, the decreasing slope after C$_2$ is much higher at 1\% \ce{CH4} than at 10\%, and higher amounts of \ce{CH4} lead to larger \textit{wings} with respect to the INMS spectra. As the neutral and ion chemistry are tightly coupled, comparing these two sets of data can shed light on protonation-driven chemical pathways, if any.\ 
We will first focus on the very light ions of the C$_1$ group. The two major C$_1$ ions are represented by \textit{m/z} 15 and \textit{m/z} 18 (Figure \ref{fig:1-5-10pc-C1}), respectively. The production of \ce{CH3+} at \textit{m/z} 15 comes from the direct ionization of methane or by charge transfer with \ce{N2+} with a branching ratio (\textit{br}) of 0.89 (Reaction \ref{CH3+formation}), as proposed by \citet{Carrasco2008}. \ce{CH3+} can react with other neutrals (Reaction \ref{CH4destruction}), and thus decreases with higher methane mixing ratios.\\

\begin{equation}
\label{CH3+formation}
  \ce{N_2^+ + CH_4 \longrightarrow CH_3^+ + H + N_2}
\end{equation}

\begin{equation}
\label{CH4destruction}
  \ce{CH_3^+ + CH_4 \longrightarrow C_2H_5^+ + H_2}
\end{equation}

\begin{equation}
\label{NH4+formation}
  \ce{NH3 + HCNH+ \longrightarrow NH4+ + HCN}
\end{equation}

\begin{equation}
\label{NH4+formation2}
  \ce{NH3 + C2H5+ \longrightarrow NH4+ + C2H4}
\end{equation}\\

In our chamber, \ce{CH3+} might form according to Reaction \ref{CH3+formation}, the same way the ammonium ion \ce{NH4+} can form via Reactions \ref{NH4+formation} or \ref{NH4+formation2} by proton attachment \citep{Vuitton2007,Carrasco2008}. Initially, \citet{Keller1998} predicted the peak at \textit{m/z} 18 to be dominated by \ce{H2O+}. After the early T5 Titan flyby, \citet{Cravens2006} and \citet{Vuitton2006a} first suggested \ce{NH4+} to be a good candidate for the detected peak at \textit{m/z} 18, which was confirmed by \citet{Vuitton2007} and \citet{Carrasco2008}. 
Note that in our experiment, a small \ce{H2O+} contribution from residual air in the chamber cannot be ruled out, given the \textit{m/z} 19 peak detection at $<1\%$ (Figure \ref{fig:1-5-10pc-C1} and Figure \ref{Pie charts 1,5,10pc}).
At 1\% and 5\% \ce{CH4}, \ce{NH4+} is a major precursor, representing 11\% and 13\% in intensity (Figure \ref{Pie charts 1,5,10pc}), respectively. Reactions \ref{NH4+formation} and \ref{NH4+formation2} require an ammonia reservoir. Pathways for \ce{NH3} production in our chamber were detailed in \citet{Carrasco2012}. The formation of ammonia (Reaction \ref{ammonia formation}) relies on an \ce{NH} radical reservoir, which can come from (i) radical chemistry through \ce{N + H -> NH}, or (ii) ion-neutral chemistry with Reaction \ref{NH-radical-formation}, and its subsequent recombination, Reaction \ref{NH-radical-formation2}. Note however that in the reaction volume (i.e. the plasma volume), wall effects can act as a catalyzer to adsorbed N and H, and contribute to Reaction \ref{ammonia formation}, as described in \citet{Touvelle1987,Carrasco2012}.\\

\begin{equation}
\label{ammonia formation}
  \ce{NH + H2 -> NH3}
\end{equation}

\begin{equation}
\label{NH-radical-formation}
  \ce{N2+ + CH4 -> N2H+ + CH3}
\end{equation}

\begin{equation}
\label{NH-radical-formation2}
  \ce{N2H+ + e- -> NH + N}
\end{equation}\\

When increasing the methane concentration, ammonia production also increases, thanks to the available NH radicals. However, primary aliphatics also become more abundant. So, at 10\% \ce{CH4}, \ce{NH4+} no longer becomes significant, representing only 2\% in intensity (Figure \ref{Pie charts 1,5,10pc}). The \ce{NH4+} contribution at 5\% \ce{CH4} is still important (13\%), notwithstanding the peak at \textit{m/z} 15 already dominating the C$_1$ group. At least at 1\% and 5\% \ce{CH4}, the ammonium ion \ce{NH4+} is a major gas phase precursor, and its formation couples ion and neutral chemistry. This ion seems to be a relevant primary precursor for the formation of tholins with a 5\% \ce{CH4} mixing ratio \citep{Sciamma-OBrien2010}. Further studies should explore intermediate conditions between 1\% and 5\% and examine the evolution of \ce{NH4+}\\

As seen in the previous section, the C$_2$ group prevails in all spectra.
Comparing the neutral and ion mass spectra (Figures \ref{fig:neutrals-ions-comparison-1pc} and \ref{fig:neutrals-ions-comparison-10pc}) gives us a clear evolution of the peaks. In the neutral spectrum at 1\% \ce{CH4}, the ratio between signal intensities at \textit{m/z} 28 and at \textit{m/z} 29 was $\sim$ 2 orders of magnitude. In positive ion mode, \textit{m/z$_{28/29}$} $\sim$ 82\% (Figure \ref{Pie charts 1,5,10pc}). Mass 28 can be attributed to \ce{HCNH+}, \ce{C2H4+} or \ce{N2+}. \citet{Carrasco2012} showed that an increase in the peak at \textit{m/z} 28 for neutral species with increasing \ce{CH4} initial concentration was not necessarily due to \ce{N2}, but mainly \ce{C2H4}. The formation of \ce{C2H4+} depends on methane, and occurs through the rearrangement reaction as given in \cite{Carrasco2008}:\\

\begin{equation}
\label{C2H4+ formation}
  \ce{CH2+ + CH4 -> C2H4+ + H2}
\end{equation}\\

However, according to \citet{Vuitton2007}, \ce{HCNH+} had a density over an order of magnitude greater than \ce{C2H4+} in Titan's ionosphere. Therefore, \ce{HCNH+} is likely a prevailing contributing ion at this mass in our plasma conditions, too. Further modeling of the plasma conditions will be needed in order to evaluate contributions of these species.

\ce{HCNH+} is mainly formed by proton attachment, by Reaction \ref{HCNH+ formation}, but can also directly depend on a reaction with methane (Reaction \ref{HCNH+ formation2}),\\

\begin{equation}
\label{HCNH+ formation}
  \ce{C2H5+ + HCN -> HCNH+ + C2H4}
\end{equation}

\begin{equation}
\label{HCNH+ formation2}
  \ce{N+ + CH4 -> HCNH+ + H2}
\end{equation}\\

Mass 29 can be \ce{C2H5+} or \ce{N2H+}. As noted previously, \ce{N2H+} might be dominant at 1\% \ce{CH4}, whereas \ce{C2H5+} is the main ion at higher initial methane concentrations. \citet{Yelle2010} propose a mechanism where \ce{N+} reacts with hydrogen by proton transfer (Reaction \ref{NH+ formation}), and the formed \ce{NH+} reacts with neutral \ce{N2} (Reaction \ref{N2H+ formation}).\\

\begin{equation}
\label{NH+ formation}
  \ce{N+ + H2 -> NH+ + H}
\end{equation}

\begin{equation}
\label{N2H+ formation}
  \ce{NH+ + N2 -> N2H+ + N}
\end{equation}\\

Two possible reactions can form \ce{C2H5+}, according to \citet{Vuitton2007} and \citet{Carrasco2008}:\\

\begin{equation}
\label{C2H5+ formation}
  \ce{CH3+ + CH4 -> C2H5+ + H2}
\end{equation}

\begin{equation}
\label{C2H5+ formation2}
  \ce{CH5+ + C2H4 -> C2H5+ + CH4}
\end{equation}\\

Regarding the formation of \ce{H2} relevant to Reaction \ref{NH+ formation}, \citet{Lebonnois2003}, and later \citet{Krasnopolsky2009} presented a scheme where \ce{H2} is linked to the amount of methane following \ce{H + CH2 -> H2 + CH}. 
Ongoing work in our laboratory focuses on \ce{N2}:\ce{H2} mixtures, avoiding the carbon contribution from methane. We find that \ce{N2H+} is the dominant ion for \ce{H2} mixing ratios $>$5\% (A. Chatain, private communication). So, given an abundant enough \ce{H2} reservoir, \ce{N2H+} should not be neglected compared to \ce{C2H5+}. \ce{N2H+} is therefore most likely more abundant relative to \ce{C2H5+} in \ce{CH4}-poor conditions. With increasing \ce{CH4} concentration, \ce{C2H5+} production is largely facilitated thanks to \ce{CH4} and other hydrocarbons (Reactions \ref{C2H5+ formation} and \ref{C2H5+ formation2}), although \ce{N2H+} might still be an important precursor at this mass (even with \ce{H2} $>$5 \% without the presence of carbon in the mixture).

The C$_3$ group is mainly represented by the peak at \textit{m/z} 42 (e.g. $\sim$10\% at 5\% \ce{CH4}, see Figure \ref{Pie charts 1,5,10pc}), and can be attributed to \ce{CH3CNH+} in agreement with the neutral acetonitrile already detected in abundance in \cite{Gautier2011}. In our case, the peak at \textit{m/z} 42 is largely the main peak detected. It can follow two protonation reactions:\\

\begin{equation}
\label{CH3CNH+ formation R1}
  \ce{HCNH+ + CH3CN -> CH3CNH+ + HCN}
\end{equation}

\begin{equation}
\label{CH3CNH+ formation R2}
  \ce{C2H5+ + CH3CN -> CH3CNH+ + C2H4}
\end{equation}\\

\citet{Sciamma-OBrien2014} examined the first products formed by \ce{N2}:\ce{CH4} mixtures. They also find their highest \textit{m/z} 42 intensity. Note that \ce{C3H6+} is also likely present at this mass. In another similar study by \citet{Thissen2009}, analogous conditions to Titan's ionosphere were simulated using synchrotron radiation of a small gas cell. In that study, no more than C$_2$ ions were effectively detected in N$_2$-CH$_4$ mixtures. After adding hydrocarbons such as \ce{C2H2} and \ce{C2H4}, only then were some C$_3$ and C$_4$ observed. Depending on the mixture, \ce{HCNH+}, \ce{C2H5+} and \ce{N2H+} were dominant C$_2$ precursors, confirming experimentally the importance of C$_2$ unsaturated hydrocarbons to promote ion growth, in agreement with observations of Titan's large ions by the Cassini CAPS instrument \citep{Westlake2014}. Depending on the mixture \ce{HCNH+}, \ce{C2H5+} and \ce{N2H+} were dominant C$_2$ precursors.

Finally, the most abundant C$_4$ ion is the one at \textit{m/z} 52, which increases with increasing methane concentration. \citet{Sciamma-OBrien2014} noticed an increase of this ion only when injecting \ce{C6H6} to their 10\% \ce{N2}:\ce{CH4} mixture, and attributed it to the \ce{C4H4+} benzene fragment. In our case, we detect it in all three methane conditions. Benzene may be a hydrocarbon aromatic product in our plasma, although its detection has been inconclusive \citep{Gautier2011,Carrasco2012}. Therefore, a \ce{C4H4+} attribution is ambiguous here. The cyanogen \ce{C2N2+} and propiolonitrile \ce{HC3NH+} cations are the two other potential candidates. \citet{Gautier2011} noted the presence of neutral cyanogen using gas chromatography coupled with mass spectrometry, especially at 1\% \ce{CH4}. \citet{Carrasco2012} firmly confirmed its presence. \ce{C2N2} can form from two recombining CN radicals, or from the reaction between HCN and a CN radical \citep{Yung1987,Seki1996}, and can be an important ion in particular, at 1\% \ce{CH4}. Protonated propiolonitrile \ce{HC3NH+} forms via two proton transfer reactions (Reactions \ref{HC3NH+ formation R1} and \ref{HC3NH+ formation R2}), and are consistent with higher methane amounts, and a shift in peaks from \textit{m/z} 51 to \textit{m/z} 52 (Figure \ref{fig:neutrals-ions-comparison-10pc}). This absence of peak shift in the 1\% \ce{CH4} neutral and ion spectra (Figure \ref{fig:neutrals-ions-comparison-1pc}) and \textit{m/z} 52 being the most intense to begin with, confirms the attribution of \ce{C2N2+} for \textit{m/z} 52 in the positive ion spectrum at 1\% \ce{CH4}.\\

\begin{equation}
\label{HC3NH+ formation R1}
  \ce{HCNH+ + HC3N -> HC3NH+ + HCN}
\end{equation}

\begin{equation}
\label{HC3NH+ formation R2}
  \ce{C2H5+ + HC3N -> HC3NH+ + C2H4}
\end{equation}\\

\subsection{Contribution of aliphatic, amine and nitrile positive ion precursors for tholin growth}\

The gas phase precursors present in the plasma form an extensive population of cations, whose ion-molecule reactions appear to mainly rely on protonation processes. These ions consist of nitriles (\ce{HCNH+}, \ce{C2N2+}), amine (\ce{NH4+}), aliphatics (e.g. \ce{CH5+,C2H3+,C2H5+,C3H3+}) and some imines (e.g. \ce{CH2NH2+}). Constraining the major ion tholin precursors is critical in understanding pathways to their formation in the laboratory, as analogs of Titan's aerosols.\\

Infrared absorption of film solid phase material produced in the PAMPRE reactor was performed by \citet{Gautier2012}. They found that changing the methane mixing ratio from 1\% to 10\% significantly impacted the absorption coefficient of tholins, particularly that of primary and secondary amine, aliphatic methyl and \ce{-CN/-NC} regions. At low \ce{CH4} concentration, the amine signature is most intense while the aliphatic bands are hardly present. The saturated and unsaturated nitriles are abundant, and the saturated ones mostly remain up to 5\% \ce{CH4}. This amine influence at 1\% and 5\% \ce{CH4}, mostly represented by \ce{NH4+} in the current gas phase study, is consistent with the solid phase infrared analysis by \citet{Gautier2012}. Future experiments probing intermediate methane conditions would be interesting in order to evaluate the sensitivity of main ions like \ce{CH3+}, \ce{NH4+}, \ce{HCNH+}, \ce{N2H+} to a broader range of methane mixing ratios.
Aliphatic cations are especially important at 10\% \ce{CH4}, where we detect the most ions with wider block distributions. This is in agreement with neutral gas phase studies \citet{Gautier2011,Carrasco2012,Dubois2019a} that found an increased production of volatiles with high methane concentrations. Cation pathways that might occur in these conditions rely on the efficient production of e.g. \ce{CH3+}, \ce{C2H2+}, \ce{C2H4+}, \ce{C2H5+} (see Section \ref{Chemical pathways}). On the contrary, a small methane concentration was shown by \citet{Carrasco2012} to produce almost exclusively nitrogen-bearing volatile compounds. Our results show 1\% \ce{CH4} spectra largely dominated by the two \textit{m/z} 28 and \textit{m/z} 29 C$_2$ species, attributed to \ce{N2+}/\ce{HCNH+} and \ce{N2H+}, respectively. As a result, ion-molecule reaction schemes for the production of tholins involving species such as \ce{NH4+}, \ce{N2H+} and \ce{HCNH+} might be favored.\\

\citet{Sciamma-OBrien2010} found an optimum tholin production for a 3-5\% initial methane concentration. At 5\% \ce{CH4}, the C$_2$ represents $\sim$44\% of all species, and are mainly supported by \ce{HCNH+} and \ce{C2H5+}. Our results agree with previous studies by \citet{Pernot2010,Gautier2014,Maillard2018} which indicated copolymeric HCN/\ce{C2H4} and poly-(\ce{CH2})$_m$(HCN)$_n$ structure found in the tholin material. These patterns also seem to be found as cation precursors.

The contribution of \textit{m/z} 42 attributed to \ce{CH3CNH+} is more intense than some of the C$_1$ and C$_2$ species. As seen in the previous section, its production is favored by increasing methane, specifically by \ce{HCNH+} and \ce{C2H5+}. Protonated acetonitrile could be a key positive ion precursor in the peak tholin production region of 5\% \ce{CH4}. 
A model where aliphatic chain precursors, along with some amine and nitrile cations are significant at 5\% \ce{CH4}, is consistent with the solid phase analysis by \citet{Gautier2012}.

\subsection{Comparisons with the INMS T40 measurements and group patterns}\


One of the Ion and Neutral Mass Spectrometer's (INMS) goals was to probe and analyze the neutral and ion chemistry of Titan's upper atmosphere, and thus characterize the gas composition \citep[for a more detailed review on the scientific objectives of INMS, see][]{Waite2004a}). It was also possible to determine the gas number density thanks to the closed source mode. The mass range of INMS was from 1 u to 99 u, while the energy range was from 0 eV to 100 eV and a mass resolution of $\frac{m}{\Delta m}=100$ at 10\% of mass peak height. This instrument consisted of two mechanical subsystems, (i) the closed ion source mode and (ii) the open ion source mode. The open ion source mode was used to analyze reactive neutral species and positive ions. The ions were guided through four focusing lenses and then through a set of four quadrupole switching lenses (QL). These switching lenses deflected the ions by 90º and subsequently directed them to a quadrupole mass analyzer. The QL essentially acted as an energy filter in open source mode, similar to our energy filter plates (\ref{Fig3}). Thus, it was possible to select specific masses (with their specific kinetic energies) and determine their ion energy distributions (range of 1--100 eV).\\

\begin{figure}
\centering
    
    \begin{subfigure}[b]{0.75\textwidth}
   \includegraphics[width=1\linewidth]{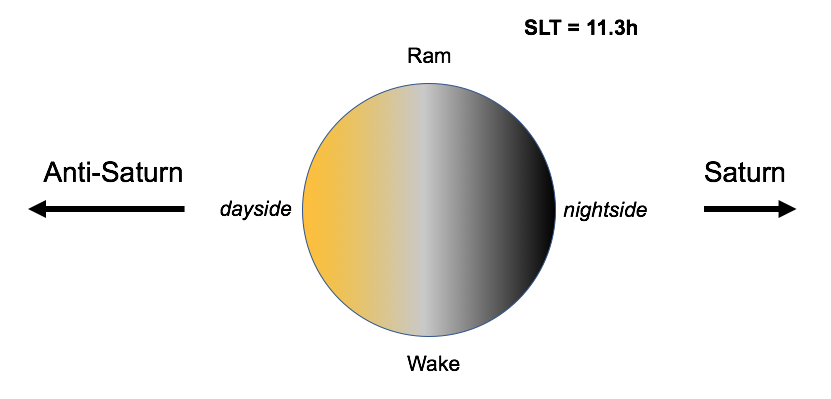}
   \caption{}
   \label{fig:sub1} 
\end{subfigure}

\begin{subfigure}[b]{0.75\textwidth}
   \includegraphics[width=1\linewidth]{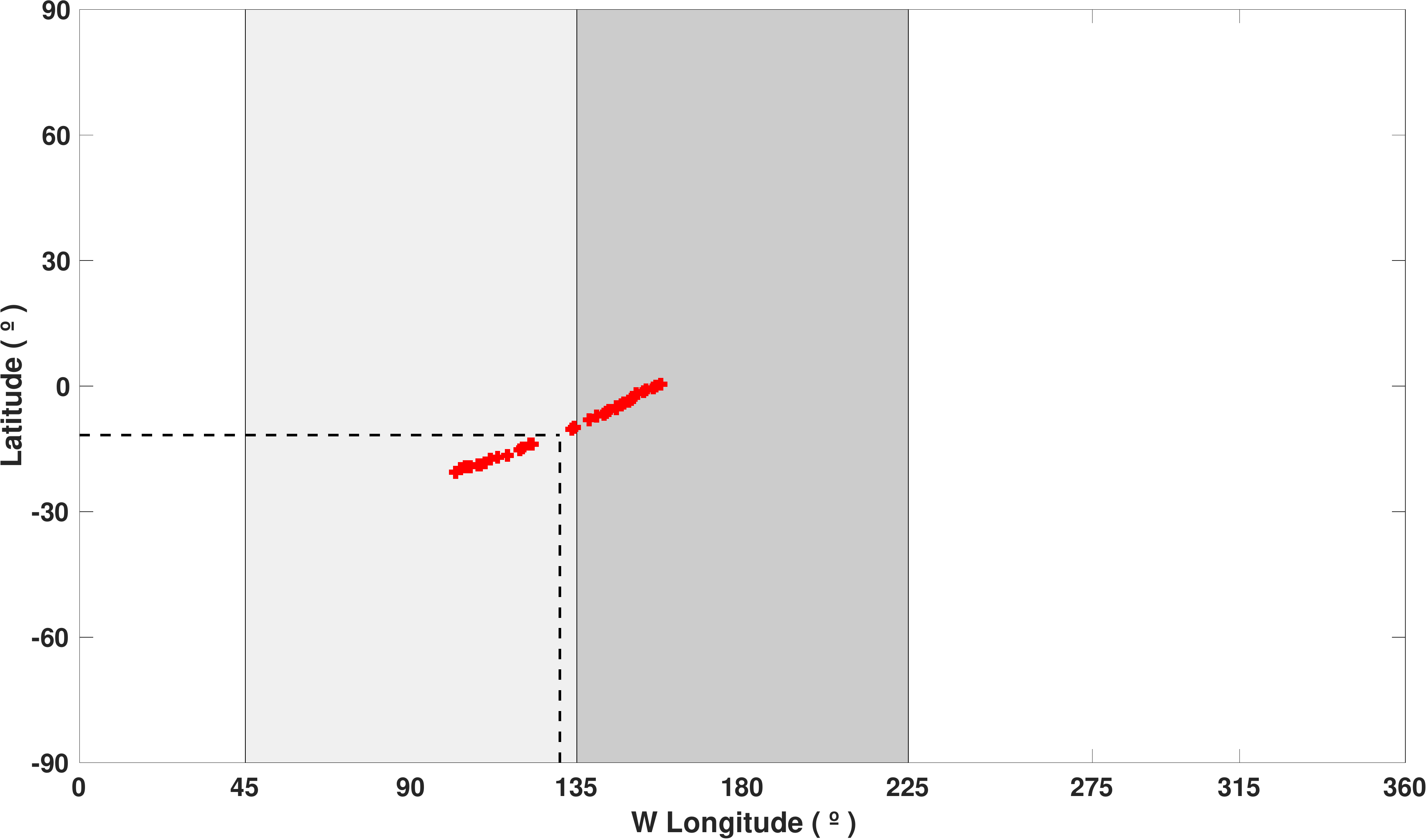}
   \caption{}
   \label{fig:sub22}
\end{subfigure}
 \caption{\label{fig:T40_sketch}(a) Titan, solar and plasma configuration during closest approach of the T40 dayside flyby on January 5, 2008. Closest approach occurred at 1010 km, at 11.7 S latitude, 130.4 W longitude, at 11.3h Saturn Local Time (SLT). (b) Flyby geometry (red) adapted from \citep{Westlake2011}. Wake region is represented in light gray and anti-Saturn region in darker gray. The dashed line represents closest approach.}
\end{figure}

We compared our laboratory spectra with positive ion measurements taken by Cassini-INMS during the T40 flyby. This encounter with Titan took place on January 5, 2008 at 13.0 h, just after Saturn local noon (sunward) with a closest approach at 1010 km above the surface (Figure \ref{fig:T40_sketch}). The passage occurred at 11.7 S latitude and 130.4 W longitude at closest approach. Cassini had a speed of 6.3 $km . s^{-1}$, with a 37.6$^{\circ}$ solar phase angle.  
During this flyby, four other instruments (VIMS, UVIS, ISS and CIRS) were also in operation. The INMS operated in open source ion mode, enabling a detection of positive ions with energies $<$ 100 eV. The integration period was $\sim 34$ ms, which corresponds to the sum of the set up/read out cycle and the sample integration period (\cite{Waite2004}). Therefore, it took INMS $\sim 2.3$ s to acquire an entire scan.\\

The fact that this flyby was entirely on the dayside of Titan and measured an important quantity of ions makes it a valuable source for the investigation of positive ions, as well as comparing it with plasma discharge laboratory simulations like PAMPRE. Solar radiation and solar photoionization was thus dominant. Furthermore, \cite{Magee2009} derived a globally averaged methane mixing ratio between 1000 and 1100 km of $\sim$ 2.17 \% which T40 was part of. The steady-state methane concentrations used in this study (from $\sim$ 0.5 \% to 5 \%) fall into or near this value \citep{Sciamma-OBrien2010}.
Previous studies have specifically focused on this T40 flyby, which, incidentally, occurred during a solar minimum activity \citep{Lavvas2011a,Richard2015a} with an $F_{10.7}$ index of 77.1. \cite{Westlake2011} found that the neutral average temperature during this flyby was 141 K. The plasma environment surrounding Titan during this encounter was unique. The measured electron distribution was categorized as bi-modal by \cite{Rymer2009}, using CAPS Electron Spectrometer (ELS) and MIMI Low Energy Magnetospheric Measurements System (LEMMS) data. They proposed that this unusual distribution, also observed at T26, T31, T46 and T47, may have been the result of two sources: the more energetic ($\lessapprox$ 5 keV) electrons coming from the plasma sheet, while the less energetic electrons ($\lessapprox$ 0.2 keV) may be local pick-up electrons from the inner magnetosphere and interactions with Enceladus water group neutrals.\\

\begin{figure}
\centering
    
\begin{subfigure}[b]{0.80\textwidth}
   \includegraphics[width=1\linewidth]{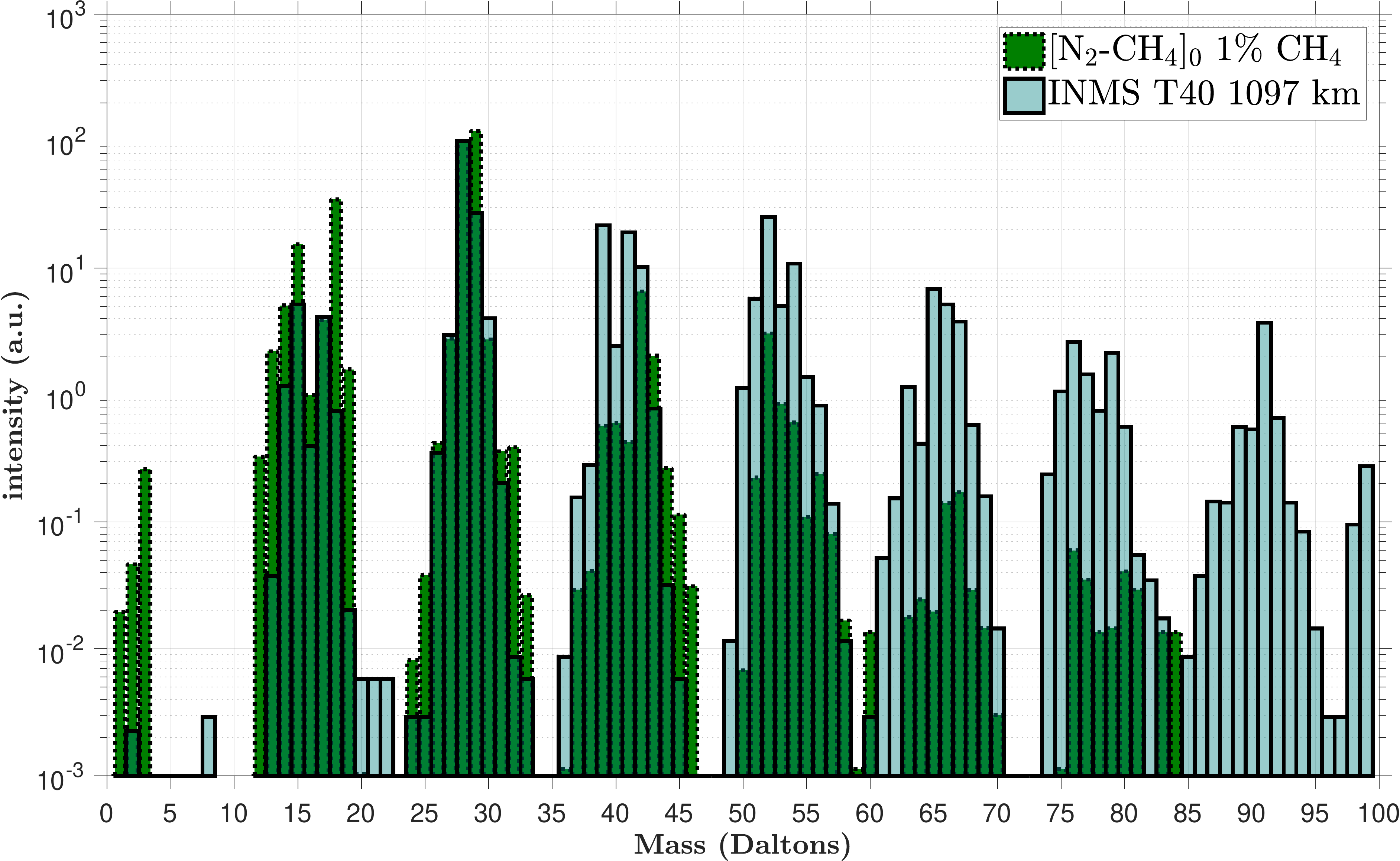}
   \caption{}
   \label{fig:sub1T40PAMPRE} 
\end{subfigure}

\begin{subfigure}[b]{0.80\textwidth}
   \includegraphics[width=1\linewidth]{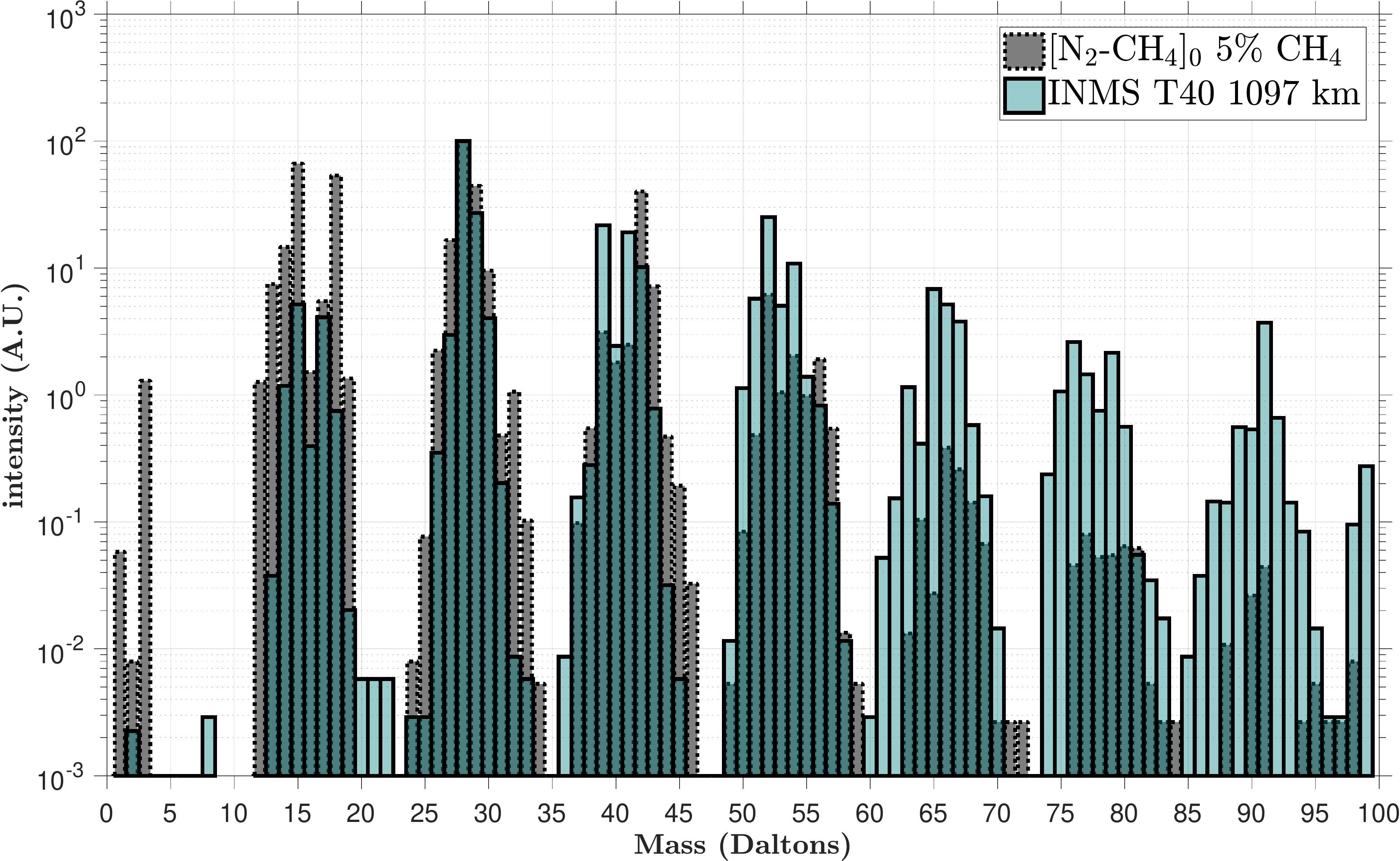}
   \caption{}
   \label{fig:sub2T40PAMPRE}
\end{subfigure}

\begin{subfigure}[b]{0.80\textwidth}
   \includegraphics[width=1\linewidth]{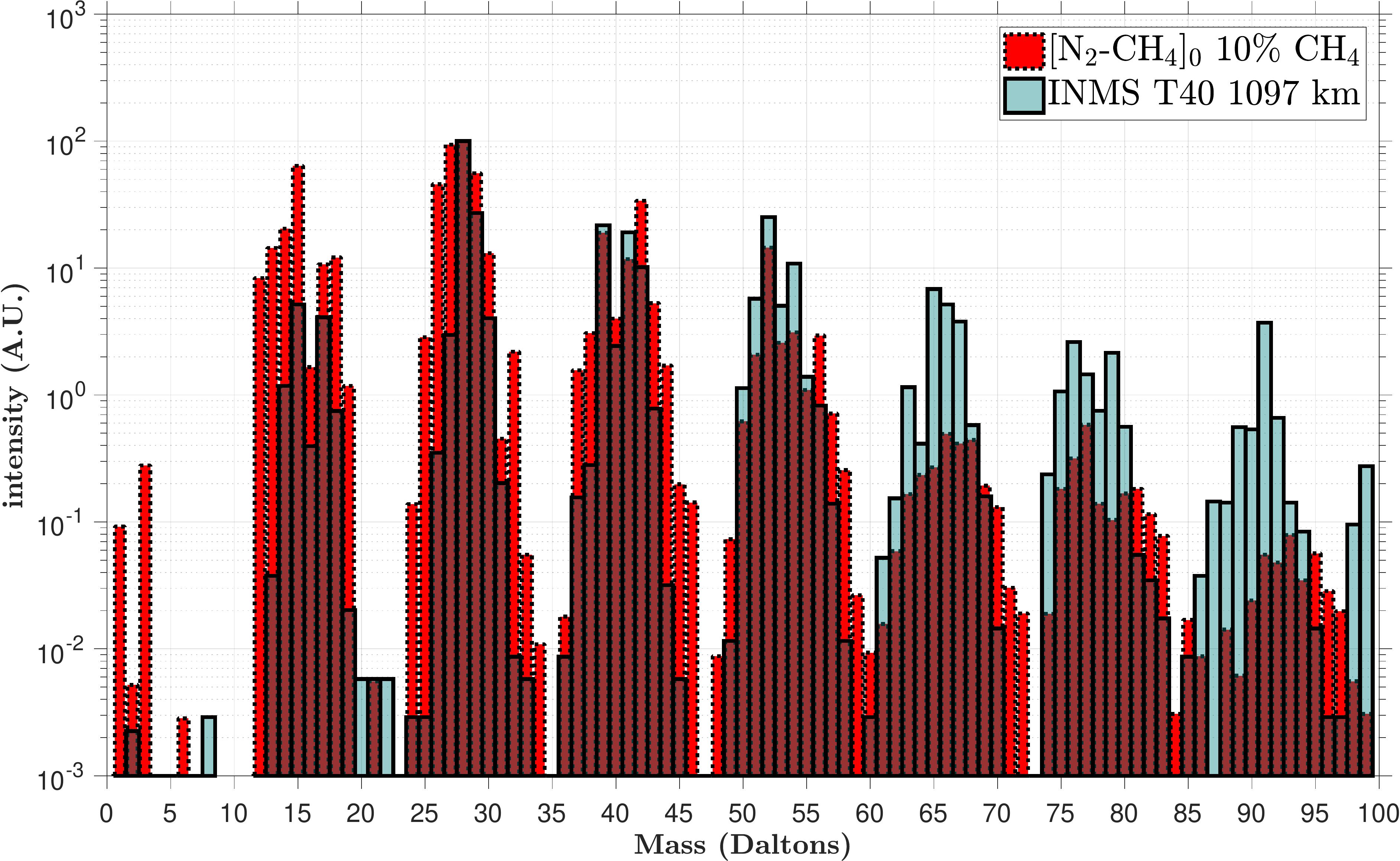}
   \caption{}
   \label{fig:sub3T40PAMPRE}
\end{subfigure}
 \caption{\label{fig:T40PAMPREcomparisons}Inter-comparison between mass spectrum taken during the outbound leg of the T40 flyby by INMS (in blue), at 1097 km, compared with our experimental averaged spectra obtained in 1\% \ce{CH4} (a), 5\% \ce{CH4} (b), and 10\% \ce{CH4} (c). The mass plots are separated in 1 u bins.}
\end{figure}

\begin{figure}
\centering
 \includegraphics[width=1.0\textwidth]{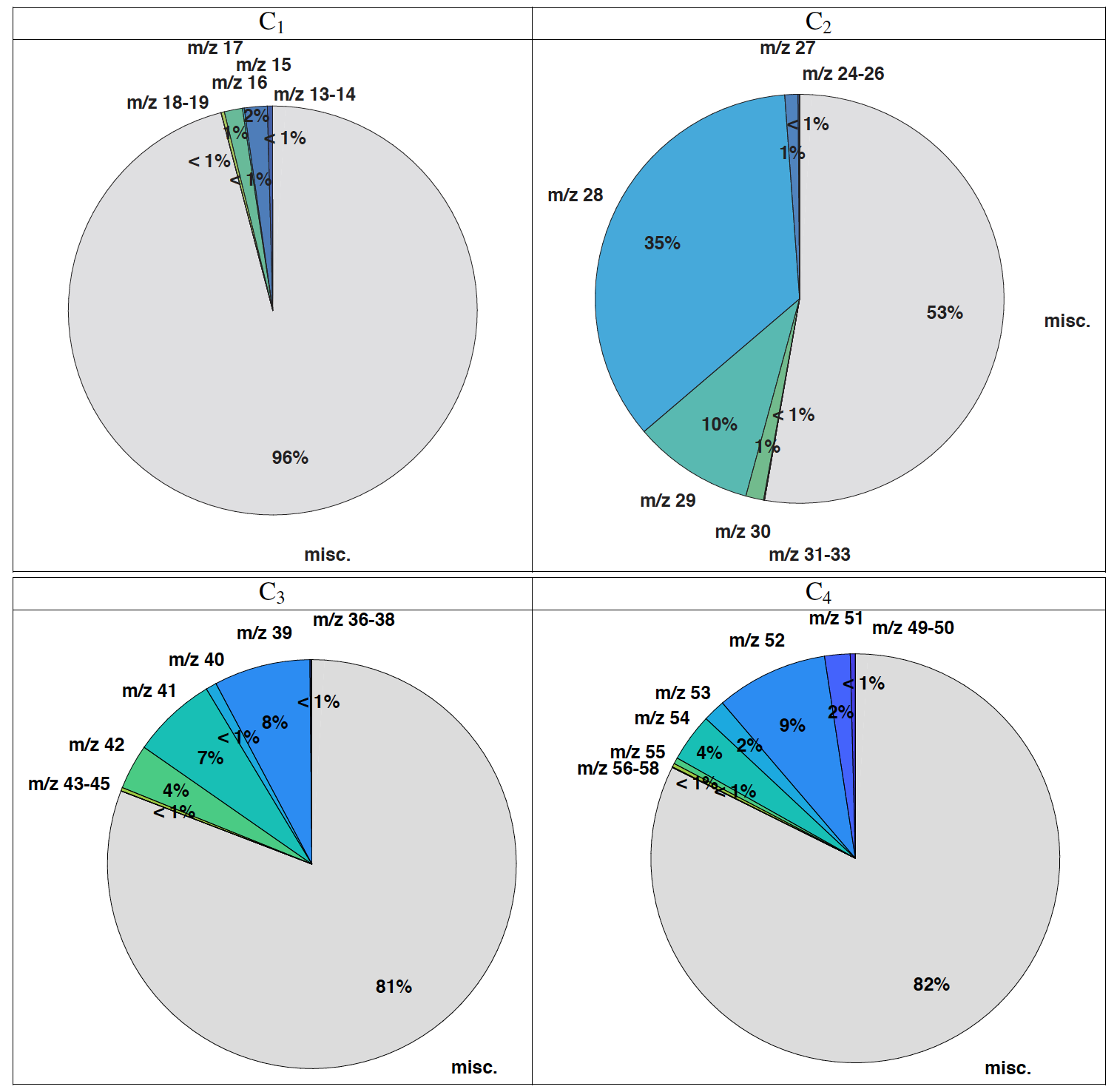}
 \caption{\label{Fig14:Pie charts T40 INMS}Pie charts of the first four C$_x$H$_y$N$_z$ groups of the T40 INMS spectrum taken at 1097 km, normalized by \textit{m/z} 28, showing the contributions of the major ions relative to the entire spectrum.}
\end{figure}

A T40 spectrum taken at an altitude of 1097 km in the outbound leg of the pass is plotted in Figure \ref{fig:T40PAMPREcomparisons}, on top of our normalized experimental 1\%, 5\% and 10\% CH$_4$ spectra (the reader is also referred to Figure 6 of the Supplementary Material for the reference spectrum). The INMS spectrum is represented as a histogram, with mass bins of 1 u. All are normalized to the intensity at \textit{m/z} 28. The T40 spectrum is made of seven clusters, each separated by about 6-8 u. The main C$_1$ compounds are at \textit{m/z} 15 and \textit{m/z} 17, which have been attributed to \ce{CH3+} and \ce{CH5+} by \cite{Vuitton2007}. The presumed \ce{HCNH+} and \ce{C2H5+} are an order of magnitude higher in intensity. Our 1\% spectrum reproduces fairly well the distributions of the grouping patterns of the lighter ions, up to the C$_4$ species (Figure \ref{fig:T40PAMPREcomparisons}a). However, the subsequent intensities decrease by orders of magnitude with each new block, with no detection beyond \textit{m/z} 84. Moreover, \ce{NH4+} is overrepresented in the experimental spectrum (11\% of the entire spectrum, Figure \ref{Pie charts 1,5,10pc}), although the efficient formation of ammonia and \ce{CH4+} may be facilitated by wall catalysis effects \citep{Touvelle1987,Carrasco2012}. Note that a water presence in the reactor with the \ce{H2O+} ion cannot be ruled out. The distribution of C$_2$ compounds is faithfully reproduced by our experimental spectra, suggesting that they at least partially play a key role in reactions with primary ions as precursors to tholin formation. As indicated in Figure \ref{Pie charts 1,5,10pc}, C$_2$ species represent $\sim$ 44\% to $\sim$ 74\% of the spectrum, depending on the \ce{CH4} condition. The equivalent charts for the T40 spectrum at 1097 km are shown in Figure \ref{Fig14:Pie charts T40 INMS}. The C$_2$ species account for $\sim$ 47\% of this spectrum, led by ions at \textit{m/z} 28 at 35\%, and \textit{m/z} 29 at 10\%. This distribution is similar to the one we find at 5\% \ce{CH4}, i.e. 25\% for \textit{m/z} 28 and 11\% for \textit{m/z} 29 (Figure \ref{fig:T40PAMPREcomparisons}b) and Figure \ref{Fig14:Pie charts T40 INMS}). The overall average C$_2$ contribution is $\sim$ 44\% in the experimental case (Figure \ref{Pie charts 1,5,10pc}), comparable with the $\sim$ 47\% in the INMS spectrum (Figure \ref{Fig14:Pie charts T40 INMS}). In the experimental spectra, increasing the methane concentration favors the production of different heavier ions, \textit{e.g.} at \textit{m/z} 39, \textit{m/z} 51, \textit{m/z} 55 and species larger than the C$_5$ group. 

The 10\% methane-rich condition (Figure \ref{fig:T40PAMPREcomparisons}c) displays the widest population of ions. We also see a rightward drift of the blocks, especially for higher masses. This is consistent with the rich presence of hydrogenated species. The major ions coincide with INMS' for the C$_1$, C$_2$ and C$_4$ groups. As seen previously, the $\frac{I_{42}}{I_{39}}$ ratio in normalized intensities reaches $\sim$ 2, when it was $\sim$ 12 with 1\% \ce{CH4}. Nonetheless, \textit{m/z} 42 remains the most abundant C$_3$ ion in all three conditions we have studied, in particular with higher methane amounts. This coincidence with the main INMS cations is particularly noticeable for masses 15, 28, 39, 41, 42 and 52. C$_x$ groups from the INMS and experimental data are asymmetrical and their respective mass ranges vary. The groups from the PAMPRE spectra obtained with 10\% of methane are wider and more species are generally detected to the right hand side of each group. These right wings are consistent with a higher contribution of saturated species in our conditions with a high methane amount.\ 
Generally, we note that fewer higher mass ions are detected in all experimental spectra (with a noticeable decrease in intensity after \textit{m/z} $\sim$ 60) compared to the INMS spectra. In the plasma discharge, the residence time of light and heavy neutral products is short (i.e. $\sim$ 28 s at 55 sccm with a pressure of 0.9 mbar) while the gas flow is fast. This lower contribution of large ions in our experiment compared to Titan's ionosphere might also be due to surface recombination of radicals occurring on the walls of the chamber, limiting their lifetime and further growth chemistry. Modeling work by \cite{Alves2012} did indeed previously show an efficient wall loss process of N radicals in the case of a pure \ce{N2} discharge.\\


The INMS measurements are characterized by a repetition of block patterns separated by 11-13 u, and 10-14 u in the PAMPRE spectra. The INMS data shows a significant contribution of the C$_2$ group, $\sim$47\% at 1097 km. \citet{Vuitton2007} determined that at the ionospheric peak during T5, the primary ions formed by ionization are quickly converted into higher mass hydrocarbons, such as \ce{C2H5+} and \ce{c-C3H3+}. The few nitrogen-bearing compounds are led by \ce{HCNH+}, \ce{CH3CNH+} and \ce{HC3NH+}, with densities of $4.6 \times 10^2$, $7.3 \times 10^1$ and $1.4 \times 10^2$ cm$^{-3}$, respectively. We find overlaps between the experimental and T40 spectra, for our 5\% and 10\% \ce{CH4} spectra, with one discrepancy at \textit{m/z} 39 and 41. The C$_3$ ions detected during T40 are dominated by the peaks at \textit{m/z} 39, \textit{m/z} 41 and \textit{m/z} 42, the latter being the most intense peak in our experimental spectra. The peak at \textit{m/z} 39 is most intense with a 10\% \ce{CH4} mixing ratio, which confirms the attribution of \ce{C3H3+}. 
Comparing these findings with neutral spectra is essential for interpretation. Acetonitrile was unambiguously attributed to \textit{m/z} 41 by \citet{Gautier2011,Carrasco2012} in their neutral gas phase studies, and correlated with methane concentration. Protonation of acetonitrile seems therefore likely to explain its significant presence, notably at 5\% \ce{CH4} (Figure \ref{Pie charts 1,5,10pc}), following Reaction \ref{CH3CNH+ formation R1}.\\

\citet{Westlake2012} used a 1-D photochemical model which was able to reproduce the ion-molecule chemistry between the primary ion products, using retrieved INMS data from the same T40 flyby. They deduced that among the two major nitrogen-bearing species in the C$_3$ group, i.e. \ce{HC2NH+} and \ce{CH3CNH+}, the latter had a higher density by a factor of $\sim$4 at the ionospheric peak. 
Overall, the 5\% experimental spectrum appears to match well the INMS spectra of the first four groups, as the intensity decreases for higher masses. The 10\% \ce{CH4} spectra also reproduce some of the higher mass features, although the group distributions present large wings, which is most likely due to unsaturated hydrocarbons. 
According to \citet{Westlake2012}, the three \textit{major} ions, \ce{CH5+}, \ce{C2H5+} and \ce{HCNH+} are connected through the following reactions:\\

\begin{equation}
\label{Major ion reaction 1}
  \ce{CH5+ + C2H4 -> C2H5+ + CH4}
\end{equation}

\begin{equation}
\label{Major ion reaction 2}
  \ce{CH5+ + HCN -> HCNH+ + CH4}
\end{equation}

\begin{equation}
\label{Major ion reaction 3}
  \ce{C2H5+ + HCN -> HCNH+ + C2H4}
\end{equation}\\

Figure \ref{Fig14:Pie charts T40 INMS} and Table \ref{key ions comparison} show a comparison of these three major ions between our experimental normalized intensities from Figure \ref{Pie charts 1,5,10pc} and their signal from the T40 flyby near the ionospheric peak ($\Sigma_{T40, 1150 km}$ row) taken from \citet{Westlake2012} ($\Sigma_{W2012}$ row). On average the 1\% and 5\% \ce{CH4} row values for these three ions match with the intensities of T40, except for \ce{C2H5+} which is much higher at 1\% \ce{CH4}. Note however that, at 1\% \ce{CH4} in particular, \ce{N2H+} may also be an important ion at \textit{m/z} 29 in our experimental spectra, as explained in Section \ref{Chemical pathways}.

\begin{table}
\centering
\caption{Relative contributions of the major ions \ce{CH5+}, \ce{C2H5+} and \ce{HCNH+} in all three initial methane concentrations, taken from Figure \ref{Pie charts 1,5,10pc}. The last row corresponds to the values from Figure \ref{Fig14:Pie charts T40 INMS}, compared with the ones from \citet{Westlake2012}, denoted W2012. The \textit{m/z} 29 contribution attributed to \ce{N2H+} at 39\% is only given in a nitrogen-rich mixture. Other values correspond to the \ce{C2H5+} attributions.}
\label{key ions comparison}
\begin{tabular}{c|ccccc}
 [\ce{CH4}]$_0$ & \ce{CH5+} & \multicolumn{2}{c}{\ce{N2H+} | \ce{C2H5+}} & \ce{HCNH+} & $\Sigma$ \\
 \hdashline 
  & \textit{m/z} 17 & \multicolumn{2}{c}{\textit{m/z} 29} & \textit{m/z} 28 &  \\
  \hline 
 1\% & 1\% & 39\% & - & 33\% & 74\% \\
 5\% & 1\% & - & 11\% & 25\% & 37\% \\
 10\% & 2\% & - & 10\% & 18\% & 30\% \\
 \hline 
$\Sigma_{W2012}$ & 2\% & - & 14\% & 50\% & 66\% \\
$\Sigma_{T40, 1097 km}$ & 1\% & - & 10\% & 35\% & 46\% \\
\end{tabular}
\end{table}

\newpage

\section{Conclusions}\

We have conducted the first positive ion analysis in our PAMPRE plasma chamber using an ion mass spectrometer, aimed at analyzing the gas phase cation precursors to the formation of Titan tholins. This setup enables us to approach the extracting tube in direct contact with the plasma and analyze the positive ions \textit{in situ}. We used three initial methane concentrations, 1\%, 5\% and 10\%, which showed a wide variability in peak distribution and size among the different conditions.\\

Furthermore, spectra taken in a 1\% \ce{CH4} mixing ratio condition are mostly dominated by nitrogen-bearing ions, such as \ce{NH4+}, \ce{HCNH+} and \ce{N2H+}/\ce{C2H5+}. We also found that the spectra using this methane concentration are largely dominated by two C$_2$ species, at \textit{m/z} 28 and \textit{m/z} 29, representing $\sim72$\% of the entire spectrum. A 1\% \ce{CH4} condition also corresponds to the most efficient gas-to-solid carbon conversion, i.e. the amount of carbon converted into the tholin product as opposed to the gas phase products \citep{Sciamma-OBrien2010}. The ammonium ion \ce{NH4+} is the third major ion with an \ce{N2}:\ce{CH4} (99:1) mixing ratio. 
Positive ions in a methane-rich mixture of 10\% \ce{CH4} are more diverse and abundant in hydrocarbon cations. Primary methane ions (\ce{CH+}, \ce{CH2+}, \ce{CH3+}) dominate the C$_1$ group. This aliphatic contribution in the gas phase agrees with aliphatic absorption present in the solid phase of 10\% \ce{CH4} tholins. The intermediate 5\% methane concentration that we chose falls in the optimum of tholin production in our plasma discharge. Here, the ion precursors are composed of protonated nitrile and aliphatic species. \ce{HCNH+} and \ce{C2H5+} are major ions. C$_2$ cations thus seem to play an important role as a gas phase precursor to tholin formation. Their dominating presence is in agreement with the HCN/\ce{C2H4} copolymerization patterns found in tholins. The largest contribution of the heavier \ce{CH3CNH+} ion is seen in this condition. \\

A majority of attributed ions seems to form via proton transfer reactions, consistent with methane injection. The primary ions of the C$_1$ group are most abundant with a 5\% \ce{CH4} mixing ratio, suggesting that at least in this condition, these species are a key precursor reservoir, available to react with larger ions. 
Our preliminary comparisons with INMS measurements performed near the ionospheric peak during the T40 flyby show promising overlaps, particularly with the 1\% and 5\% \ce{CH4} spectra. Future work should thus explore intermediate concentrations to constrain the importance of the primary ions as key precursors involved in the reaction schemes leading to the formation of Titan tholins combined with modeling work investigating the ion-neutral coupling in the \ce{N2}:\ce{CH4} discharge. Finally, while ground-based detections of cations in Titan's upper atmosphere have yet to surface, their measurements would enhance Cassini's legacy, especially since Titan passed aphelion in April of 2018 entering an 8-year southern winter, which could potentially impact the volatile distribution in the upper atmosphere. Relatively abundant species such as \ce{HCNH+}, \ce{C2H5+} or \ce{CH2NH2+}, potentially involved in hydrogenation processes with prebiotic interest, would make good targets for observations in synergy with theoretical predictions.\\

\appendix
\newpage
\section*{Acknowledgements}
The authors are indebted to two anonymous reviewers whose insightful comments substantially improved the quality of the manuscript. We acknowledge financial support from the European Research Council Starting Grant (grant agreement no. 636829), and data provided by the Cassini INMS instrument available through the PDS. We are also grateful for discussions on cation chemistry with Dr. Jérémy Bourgalais. 

\section*{Supplementary Material}

\section{Materials and methods}
\subsection{Positive ion mass spectrometry}

The multiplier potential is set at 1800
V. Within the transfer tube, in positive ion mode, two guiding lenses are set at -19 V
and -102 V, respectively.

Table \ref{Table 4.1.} lists all the intrinsic parameters pertaining to the spectrometer’s optics used for our positive ion measurements. The different components can be seen in
more detail in Figure 3. The detector consists of the multiplier (Figure 3). It is connected with (i) the 1st dynode which sets the voltage on the front of the detector and (ii) the discriminator setting a threshold on the pulse output of the multiplier. The Extraction group controls the gating system. This gating is only switched on and operational if the RF head is connected through a transistor-transistor logic (TTL) system, which was not the case in this study. The flight-focus mainly controls the guiding of the ions. The energy filter (described in the next section) is tunable along with a set of other lenses downstream of the flight-focus. The emission source is only set in neutral mode. Ion extraction was made possible by applying a negative polarity to the aperture, thus directly extracting said ions from the first chemical steps of the expanding plasma. The main relevant varying parameters here are the \textit{lens1} and \textit{extractor} potentials. The latter has a negative value for positive ion extraction. As we will see in the following section, the energy variable part of the Sector component determines the energy filtering which will be fine-tuned for each gas mixture.

\begin{table}
	\centering
	\caption{Global Environment Editor example for a given positive ion mass spectrum acquisition using the MASoft Hiden Analytics software. The Group column corresponds to the different components of the Electrostatic Quadrupole Plasma system, as shown in Figure 3. The second column represents all of the tunable variables, and their respective values, either fixed after a tune, or changed by the user.}
	\label{Table 4.1.}
	\begin{tabular}{lccc} 
		\hline
		Group & Name & Value & Units\\
		\hline
        \multirow{3}{*}{Detector} &
		1st dynode & -3500 & V\\
		& discriminator & -10 & \%\\
		& multiplier & 1800 & V\\
        \hline
        \multirow{4}{*}{Extraction} &
        extractor & -191 & V\\
        & gate delay & 0.1 & $\mu S$\\
        & gate width & 0.0 & $\mu S$\\
        & lens 1 & -19 & V\\
        \hline
        \multirow{1}{*}{Flight-focus} &
        flight-focus & -69 & V\\
        \hline
        \multirow{5}{*}{Quad} &
        delta-m & 0 & \%\\
        & focus2 & -176 & V\\
        & resolution & 0 & \%\\
        & suppressor & -200 & V\\
        & transit-energy & 0.0 & V\\
		\hline
        \multirow{7}{*}{Sector} &
        D.C. quad & 22 &\%\\
        & axis & -40.0 & V\\
        & energy & 2.2 & V\\
        & horiz & -5 & \%\\
        & lens2 & -102 & V\\
        & plates & 7.4 & V\\
        & vert & -20 & \%\\
        \hline
        \multirow{3}{*}{Source} &
        cage & 0.0 & V\\
        & electron-energy & 70.0 & V\\
        & emission & 100.0 & $\mu A$\\
        \hline
        \multirow{6}{*}{Other}&
        CRV-range & 7 & $c/s$\\
        & CRV-resolution & 6 & bits\\
        & gating & 0 & (1=on, 0=off)\\
        & gating-invert & 0 & (0=normal, 1=invert)\\
        & mass & 28.0 & u\\
        & mode-change & 1000 & ms\\
	\end{tabular}
\end{table}

\subsection{Ion filter energy profiles}

Ion energy studies in plasma discharges have previously shown their dependences to experimental setups and plasma compositions.
\cite{Field1991} first reported on the energy profile of certain ions (e.g. \ce{O2+}) derived from theoretical calculations and compared with experimental results in a radiofrequency plasma discharge. Further studies examined the role of plasma rf sheaths
and pressure-dependence on ion energy distributions \citep{Gudmundsson1999,Kawamura1999,Wang1999}.
In this setup, it is possible to measure and quantify the Energy Distribution (ED)
at the energy filtering plate, for specific masses in different plasma conditions. The
energy parameter from the Sector group (Table \ref{Table 4.1.}) corresponds to the energy filter (plates) of Figure 3 and controls the energy of the ions which are mass analyzed by the Quadrupole Mass Spectrometer (QMS) detector later on. Therefore, a lens optimization is crucial before each mass analysis. For Titan, as stated in the previous section, different ions may have different kinetic energies during detection by
INMS. This variation may be due to atmospheric composition, plasma flow composition,
spacecraft velocity, \citep{Waite2004,Lavvas2013}.
In our experiments, the main fluctuating parameter is the \ce{N2}-\ce{CH4} mixing ratio.
Corresponding mass spectra with an optimized energy filter will be presented in the following section.

\subsection{Protocol}

As stated previously, the analysis of cations in the plasma discharge requires that (i) the spectrometer not
be obstructed by incoming tholins to provide accurate intensity measurements and
not interfere with the optics potentials in the transfer tube, and (ii) the mass analysis
be optimized to the corresponding energy distribution of the highest mass (\textit{m/z} 28);
the latter being significant in most methane mixing ratios. A set of four experiments
is carried out for each chosen mixing ratio. Every other experiment, we
"calibrate" our measurement by carrying out an \ce{O2} plasma in specific conditions for a given steady distance from the plasma. This critical step in controlled conditions assures us that the extractor
is not obstructed by tholins, by setting a reference intensity value at \textit{m/z} 32 (i.e. \ce{O2+} at $\sim 10^5$ c/s and $<10^2$ c/s when the aperture is obstructed). If the intensity of the latter mass has not decreased from one mixing ratio to another, this means tholins have not accumulated on the extractor. Thus, measurements from one experiment to the next are reliable. This is an important step to take into account,
given how efficient the setup is to produce tholins \citep{Sciamma-OBrien2010}. Furthermore, energy distributions are measured through energy filter plates and optimized before each experimental run, in order to tune the filter system for mass analysis.

\section{Results}

\subsection{Energy Filter Distributions of selected species}

\begin{figure}
\centering
\includegraphics[width=0.9\textwidth]{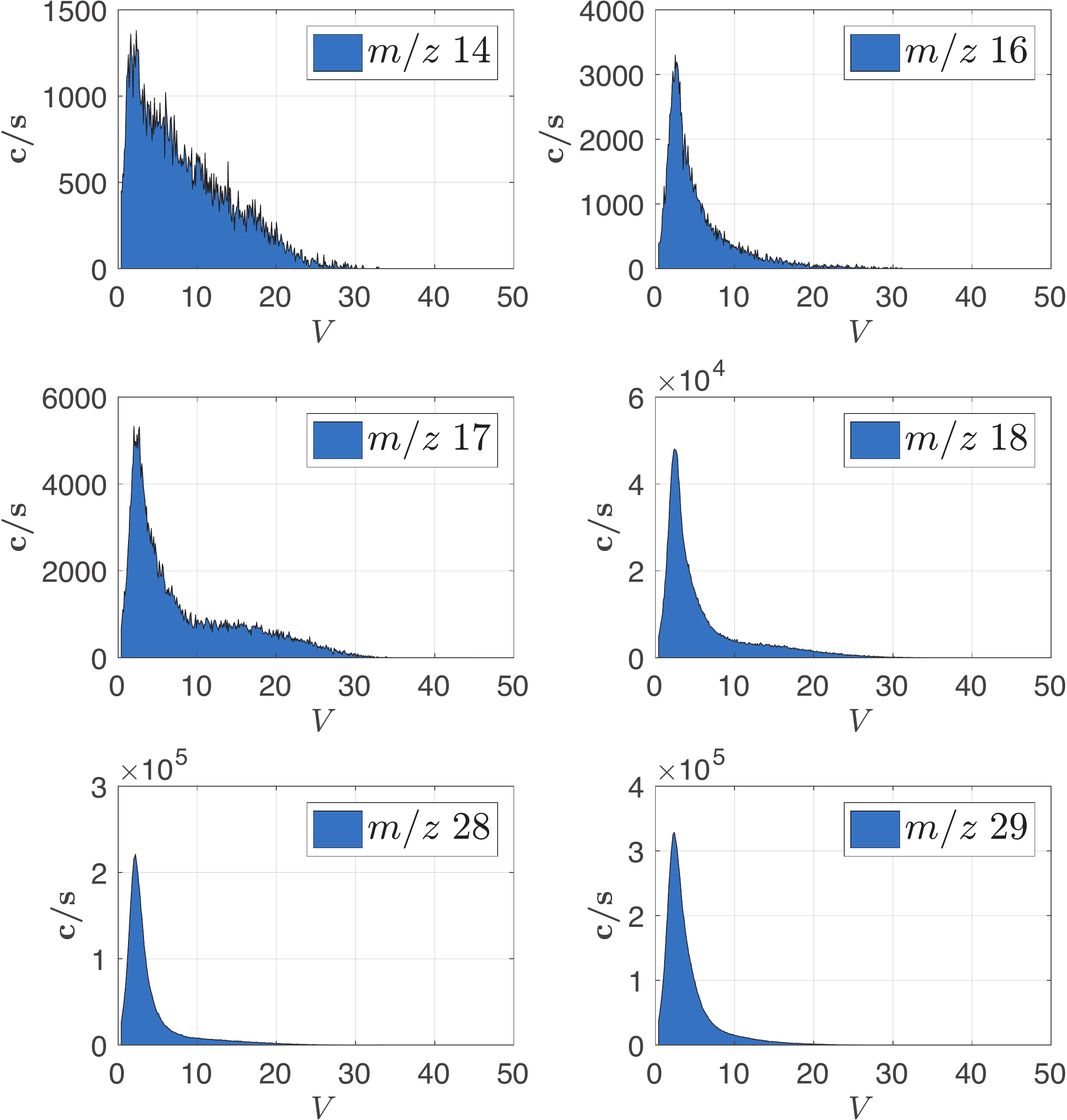}
\caption{\label{Fig4}Energy filter distributions in an [\ce{N2}-\ce{CH4}]$_0$ = 1\% mixing ratio for selected
ions \textit{m/z} 14, 16, 17, 18, 28 and 29.}
\end{figure}

\begin{figure}
\centering
\includegraphics[width=1.0\textwidth]{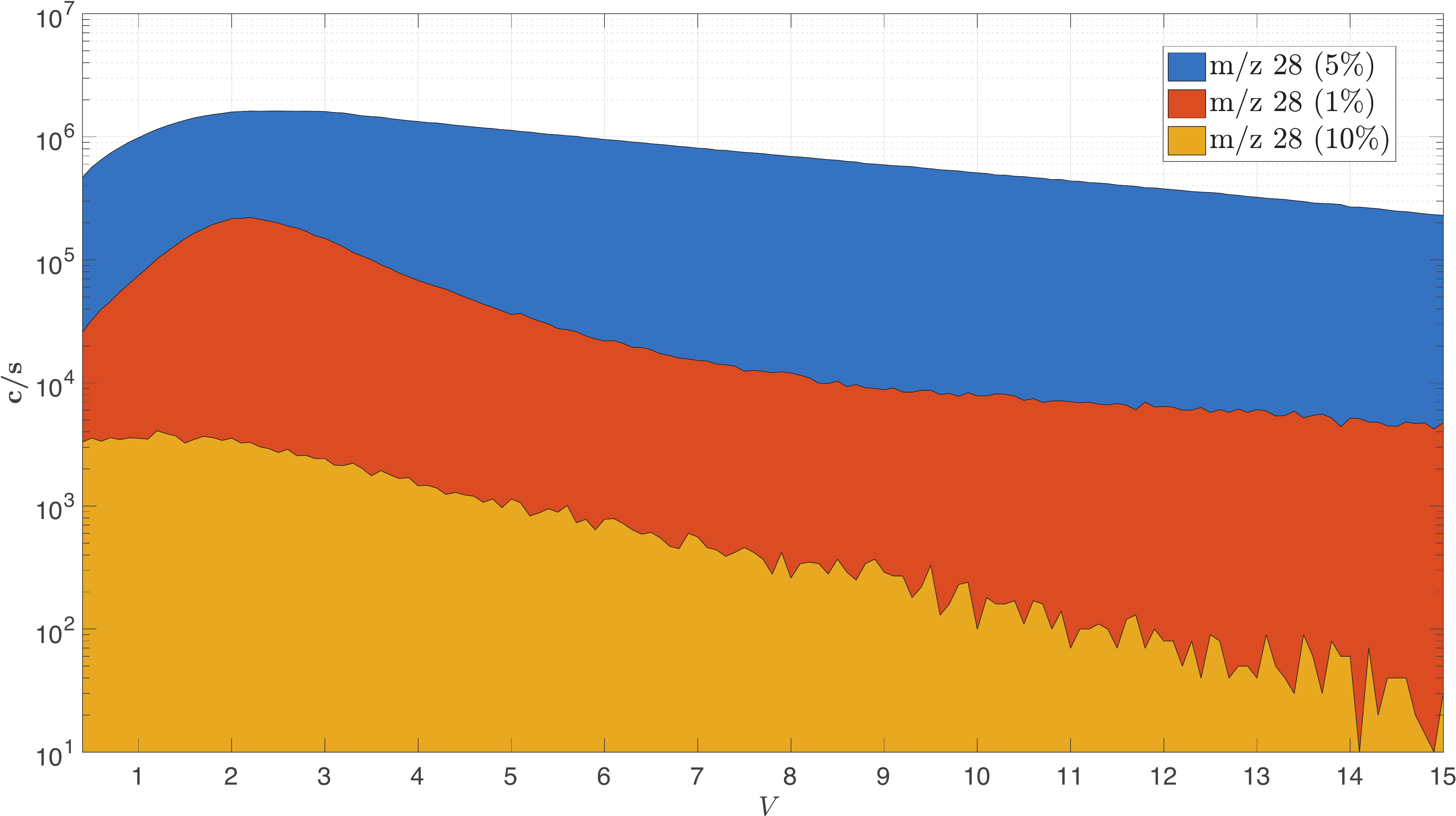}
\caption{\label{Fig5}Energy filter distributions in three [\ce{N2}-\ce{CH4}]$_0$ concentrations, 1\%, 5\% and 10\% for
ions at \textit{m/z} 28.}
\end{figure}

The energy filter (Figure 3) scans the entirety of available potentials ($<$ 100 V) for selected masses. This parameter represents the potential applied to the energy
filter plate, between the \textit{focus2} lens and the QMS. Consequently, one can obtain fairly reproducible energy distributions for different ions which are part of the same gas mixture and at the same pressure. Hence it is possible to get such a distribution
for any given plasma discharge condition. To get a good estimate of these energy distributions (EDs) and
to make sure the energy distributions for specific ions are representative of the entire
plasma, we chose six different ions: \textit{m/z} 14, 16, 17, 18, 28 and 29, which are
expected to be present and detected in our plasma. Figure \ref{Fig4} shows the EDs measured at the energy filter electrode. The \textit{m/z} 28 and \textit{m/z} 29 distributions are likely the most representative of two major compounds given their high count intensities.
Nonetheless, these EDs are consistent in their distribution for one given condition,
as well as for ions of different masses. The distribution for \textit{m/z} 14 is affected by low
intensity counts, but all other five have Maxwellian distributions, centered at 2.2 V
for 1\% \ce{CH4}, 2.2 V for 5\% \ce{CH4} and 1.2 V for 10\% \ce{CH4} (Table \ref{table:IED_maxima}). These differences imply a difference in chemistry and ion composition in the plasma. Figure \ref{Fig5} shows the ED taken at \textit{m/z} 28 for [\ce{CH4}]$_0$=1\%, 5\% and 10\% mixtures. 


EDs can sometimes be bi-modal, as can be slightly seen for \textit{m/z} 17. This bi-modal distribution could be an instrumental artifact (lens coating, re-acceleration of ions in the skimmer) or attributed to plasma sheath effects \citep{Kawamura1999}.
Figure \ref{Fig5} shows the superimposed EDs for \textit{m/z} 28 in all conditions. The tail of
the Maxwellian starts decreasing by at least an order of magnitude after $\sim$ 3 V, with peaks between 1 and 3 V. 



\begin{table}
	\centering
	\caption{Energy distribution maxima for [CH$_4$]$_0$=1\%, 5\% and 10\%.}
	\label{table:IED_maxima}
	\begin{tabular}{cc} 
		\hline
		[CH$_4$]$_0$ (\%) & energy (V)\\
		\hline
        1 & 2.2\\
        
        5 & 2.2\\
        
        10 & 1.2\\
        \hline 
	\end{tabular}
\end{table}

\begin{figure}
\centering
\includegraphics[width=1.0\textwidth]{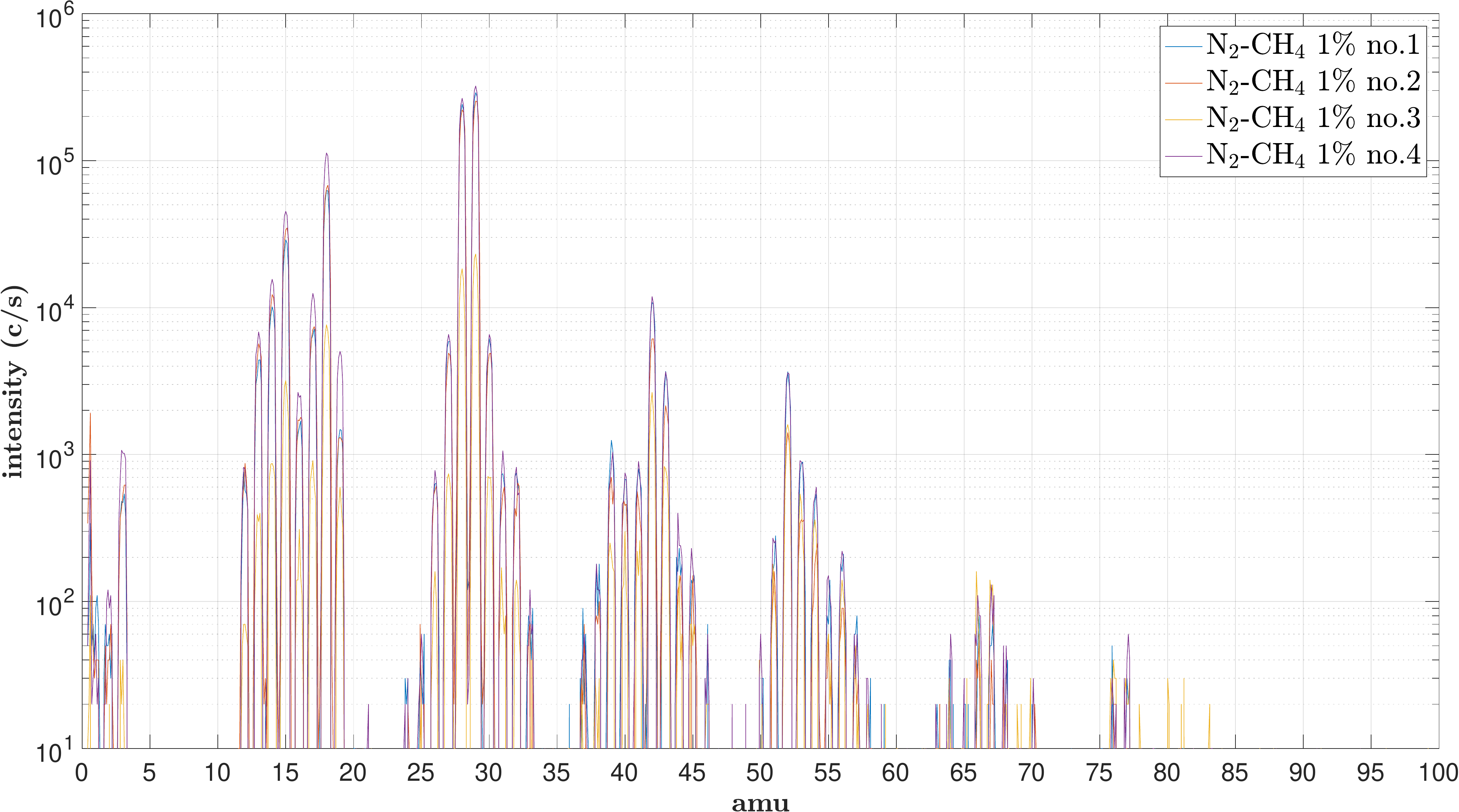}
\caption{\label{Fig6}Mass spectra for a [\ce{N2}-\ce{CH4}]$_0$ = 1\% mixing ratio, with
the positive ion energy filter set at 2.2 eV, i.e. the maximum energy
for \textit{m/z} 28 (see Figure \ref{Fig5}). Four spectra are shown here and were taken in the same experimental conditions (plasma pressure, gas mixture).}
\end{figure}


\begin{figure}
\centering
\includegraphics[width=1.0\textwidth]{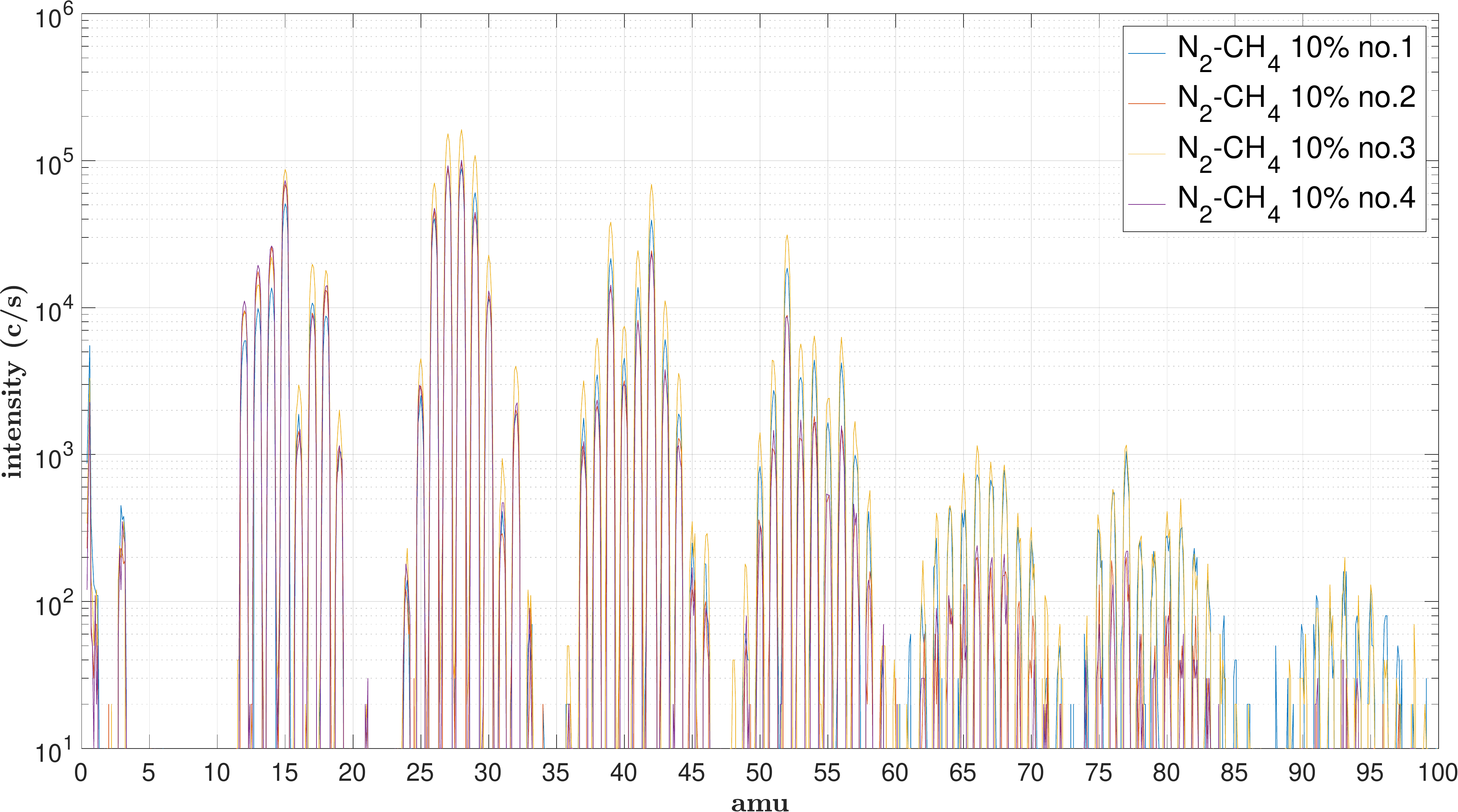}
\caption{\label{Fig7}Mass spectra for a [\ce{N2}-\ce{CH4}]$_0$ = 10\% mixing ratio,
with the the positive ion energy filter set at 1.2 V, i.e. the maximum
energy for \textit{m/z} 28 (see Figure \ref{Fig5}). Four spectra are shown here and were taken in the same experimental conditions (plasma pressure, gas mixture).}
\end{figure}

We acquired four spectra in each condition (Figures \ref{Fig6} and \ref{Fig7}). We see a temporal signal variability, attributed to the potential instabilities between the extractor and the plasma. For example, the \textit{m/z} 28 intensity at 10\% (Figure \ref{Fig7}) varies from $8.82 \times 10^4$ to $1.62 \times 10^5$ c/s, and from $8.67 \times 10^3$ to $3.12 \times 10^4$ c/s for \textit{m/z} 52 (33,958 and 10,636 standard deviation, respectively). At 1\% CH4, the \textit{m/z} 28 and \textit{m/z} 52 intensities vary from $1.83 \times 10^4$ to $2.65 \times 10^5$ c/s, and $1.41 \times 10^3$ to $3.66 \times 10^3$ c/s, respectively (standard deviations of 186,000 and 1229). Other factors such as tholin coating on the extractor or intrinsic plasma variations may also play a role in these fluctuations. This is why we will henceforth normalize all spectra at \textit{m/z} 28, and units will be expressed as arbitrary units (A.U.).

\begin{figure}
\centering
 \includegraphics[width=1.0\textwidth]{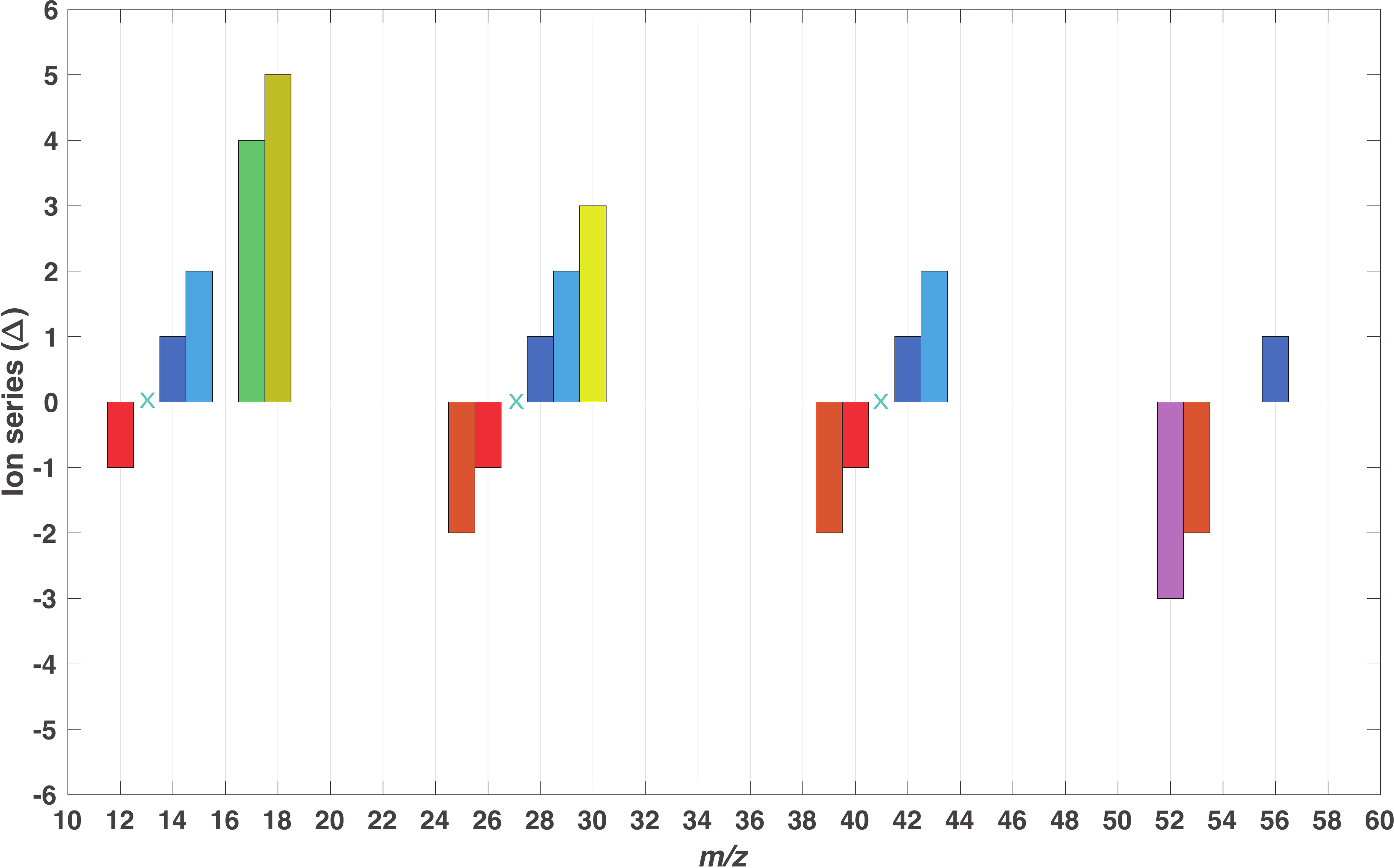}
 \caption{\label{deltavalues}Delta ($\Delta$) values for 20 selected masses covering the C$_{1-4}$ groups shown in Figure 9 and using the same color code. As per the analyses in \cite{Canagaratna2007} and \cite{McLafferty1993}, each mass of the spectra is assigned a $\Delta$ value, where $\Delta = m/z - 14n +1$ ($n$ corresponds to the grouping number). Typically, $\Delta = 2$ corresponds to linear saturated hydrocarbons, $\Delta > 2$ corresponds to heteroatomic compounds and $\Delta < 2$ corresponds to unsaturated and branched hydrocarbons. Therefore, the $\Delta$ value informs on the probable functional groups at each mass. Note that $\Delta = 0$ values are indicated as blue crosses.}
\end{figure}

\begin{figure}
\centering
 \includegraphics[width=1.0\textwidth]{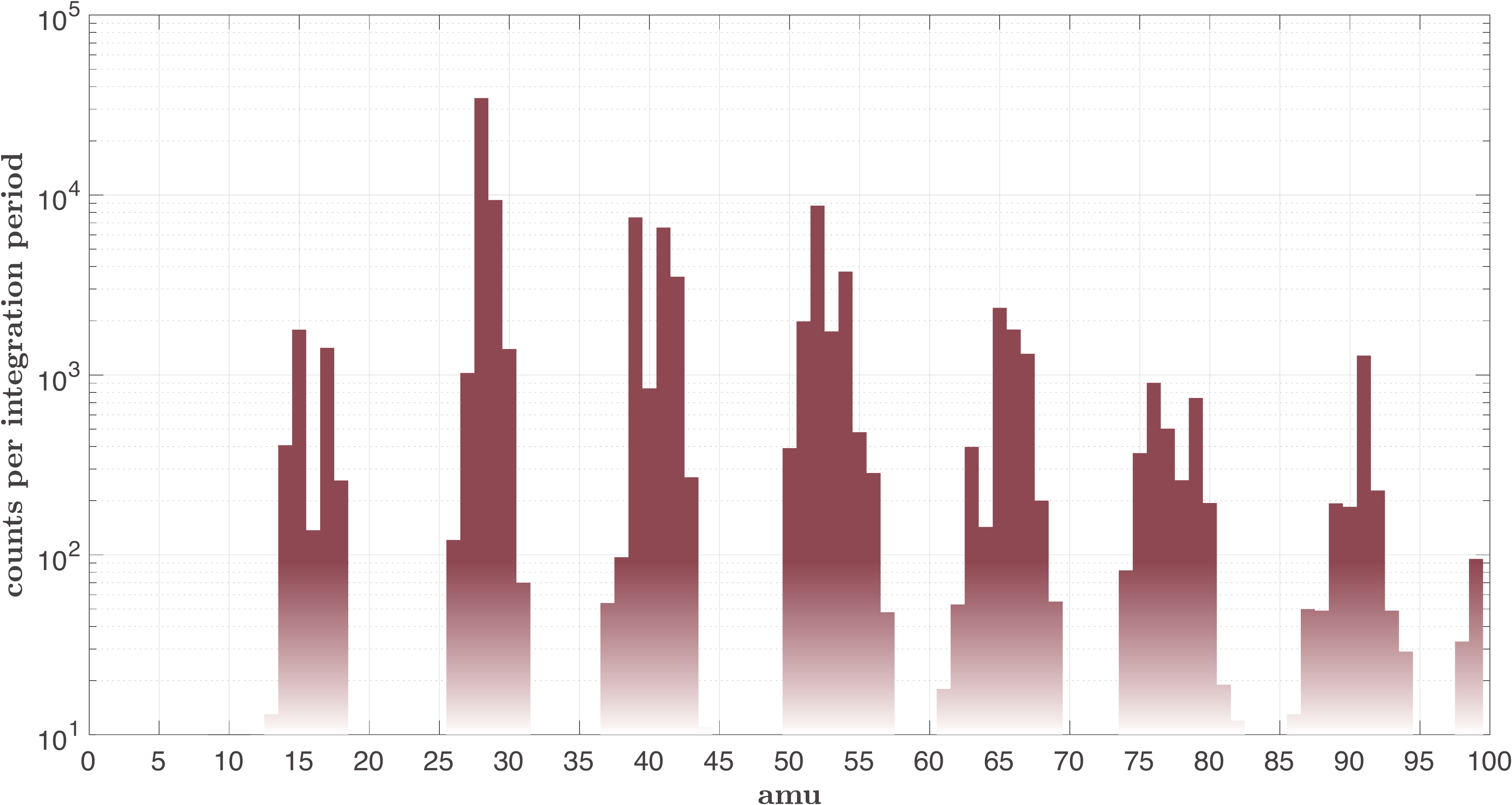}
 \caption{\label{fig:T40_1097km}Mass spectrum taken by INMS during the outbound leg of the T40 flyby, at 1097 km. The mass plot is separated in 1 u bins and plotted against raw IP counts. The INMS operated in open source ion mode during this flyby in order to detect low energy ions ($<$ 100 eV).}
\end{figure}

\newpage


 \bibliographystyle{elsarticle-harv} 
 \bibliography{Bib_July2019.bib}









\end{document}